\RequirePackage{ifpdf}
\ifpdf
\documentclass[pdftex,preprint,prb, showpacs,amsmath,amssymb,floatfix,aps ]{revtex4}
\else
\documentclass[preprint,prb, showpacs,amsmath,amssymb,floatfix,aps]{revtex4}
\fi

\usepackage{graphicx}
\usepackage{amsmath,graphics,  bm}
\usepackage{subfigure}

\begin{document}

\author{Wijnand Broer, George Palasantzas, and Jasper Knoester}
\affiliation{Zernike Institute for Advanced Materials, University of Groningen, \\
Nijenborgh 4, 9747 AG Groningen, the Netherlands}

\author{Vitaly B. Svetovoy} 
\affiliation{MESA\textsuperscript{+} Institute for Nanotechnology, University of Twente, \\
P.O. Box 217, 7500 AE Enschede, the Netherlands}
\affiliation{{Institute of Physics and Technology, Yaroslavl Branch, Russian Academy of Sciences, 150007 Yaroslavl, Russia.}}

\title{Roughness correction to the Casimir force at short {separations}:\\ Contact distance and extreme value statistics}

\date{\today}

\begin{abstract}
So far there has been no reliable method to calculate the Casimir force at separations comparable to the root-mean-square of the height fluctuations of the surfaces. Statistical analysis of rough gold samples has revealed the presence of peaks considerably higher than the root-mean-square roughness. These peaks redefine the minimum separation distance between the bodies and can be described by extreme value statistics. Here we show that the contribution of the high peaks to the Casimir force can be calculated with a pairwise additive summation, while the contribution of asperities with normal height can be evaluated perturbatively. This method provides a reliable estimate of the Casimir force at short distances, and it solves the significant, so far unexplained discrepancy between measurements of the Casimir force between rough surfaces and the results of perturbation theory. Furthermore, we illustrate the importance of our results in a technologically relevant situation.
\end{abstract}

\maketitle

\section{Introduction}
The Casimir force is an electromagnetic dispersion force of  quantum mechanical origin between neutral bodies without permanent dipoles. It  was introduced as the effect of retardation on Van der Waals forces. \cite{CasimirPolder48, Casimir48} Later, it was generalized to arbitrary dielectric plates at finite temperatures, which revealed how this force depends on the frequency dependent permittivities $\varepsilon(\omega)$ of the interacting materials.\cite{Lifshitz55, Lifshitz61} Early measurements\cite{Sparnaay58, vBlokland78} hinted at the existence of the Casimir force, whereas the first high accuracy measurements were performed only decades later with the use of a torsion pendulum.\cite{Lamoreaux97} Other techniques, such as atomic force microscopy (AFM) and micro-oscillator devices were employed later in a plate-sphere geometry \cite{Decca2005, Mohideen2006}(Fig. \ref{Fig:platesphere}). Other configurations were also investigated, e.g. parallel plates\cite{ParallelPlates2002} and crossed cylinders.\cite{CrossedCylinders2000}
\par Nowadays, electromechanical engineering is  being conducted at the micron scale, and has regenerated interest in the Casimir force because of its significance in the distance range of nanometers up to microns. Microelectromechanical systems (MEMS) have the right size for the Casimir force to play a role: their surface areas are large enough, but their gaps are small enough for the force to draw components together and possibly lock them together.\cite{Maboudian97} This effect, known as stiction, in addition to capillary adhesion due to the water layer present on almost all surfaces, is a common cause of malfunction in MEMS devices.\cite{Serry98, BuksEPL2001, BuksPRB2001} Moreover, the development of increasingly complex MEMS will attract more attention to scaling issues as this technology evolves towards nanoelectromechanical (NEM) systems. The issue of Casimir interactions between surfaces in close proximity will inevitably need to be faced, with particular attention paid to stiction due to Casimir and other surface forces. Besides stiction and associated pull-in instabilities in MEM actuation dynamics, the Casimir force can be utilized \cite{CapassoReview2007} to control actuation dynamics in smart ways, leading to development of ultrasensitive force and torque sensors that can even levitate objects above surfaces without disturbing electromagnetic interactions and without friction to translation or rotation.\cite{Johnson2011} 
Finally, from a  more fundamental viewpoint, the Casimir force plays an important role in the search for hypothetical new forces.\cite{Onofrio2006}
\par There are three effects that must be accounted for when calculating the Casimir force between real interacting surfaces: The influence of optical properties of the materials, surface roughness, and temperature contributions. Temperature has been shown to have a significant effect only at separations larger than 1 $\mu$m, because at shorter separations the thermal modes do not fit between the surfaces at room temperature.\cite{Milton2004} However, at separations less than 1 $\mu$m, especially in the range below 100 nm, the influence of optical properties and surface roughness should be carefully taken into consideration. 
\par Scattering on rough surfaces is a stochastic process: in general there is insufficient information to derive an exact roughness correction to the Casimir force. A possible way to cope with this is a perturbative approach:\cite{vanBree74, Genet2003, Lambrecht2005} it is assumed that a rough surface is a small deviation from a smooth surface. Moreover, the slopes of the surface profiles must be small. Such assumptions provide enough constraints to come to an analytical expression for the Casimir force between rough bodies. This approximation is valid at separations $d$ much larger than the root-mean-square (rms) roughness $w$: $d\gg{}w$. For $d\sim{}w$ there is no analytical solution to the problem. This is why there is no (analytical) method beyond perturbation theory.  Another method to estimate dispersion forces is the so called proximity force approximation (PFA).\cite{Derjaguin34}  When applied to rough surfaces \cite{Kli96} it assumes that the force between rough surfaces can be presented as the sum of forces between opposing flat surfaces. {This method  is valid in the case of small separations in comparison to the correlation length $\xi$: $\xi\gg{}d$}, because it assumes the contribution of different patches to be independent of each other.
\par Statistical analysis of rough gold samples has demonstrated the presence of peaks considerably higher than $w$.\cite{d0paper} In this paper, we will show that the contribution of these peaks can be calculated with the PFA, and that the contribution of the height values closer to the average can be evaluated perturbatively. This distinction gives a reliable estimate of the Casimir force at short separations. It was introduced in a recent letter,\cite{BroerEPL2011} where it was shown to reproduce experimental results \cite{PeterRoughness} for one gold sample. In the present paper, the method will be discussed in more detail, and results for multiple gold samples will be shown. Moreover, this paper includes an estimate of the influence of the shapes of the peaks, and a prediction of the Casimir force in a technologically relevant situation.
\par The paper is organized as follows: after the introduction it will provide the starting points of this approach: Lifshitz theory and the statistics of rough surfaces. This is followed by an outline of the model with derivations of the main formulas. Section \ref{curvature} will specifically address the role of the {shape} of the peaks. In section \ref{direct_bonding}, we will present a prediction for a relatively smooth sample. In such a case, force measurements are hindered by jump to contact, but force predictions are useful for applications in direct bonding technology. \cite{Haisma2007} Just before the final section with the conclusions, we will evaluate the results and compare them to experimental data from Ref. \onlinecite{PeterRoughness}.

\section{Starting points and assumptions for force calculations}
\subsection{Lifshitz theory} 
\begin{figure}
	\centering
		\includegraphics[width=0.45\textwidth]{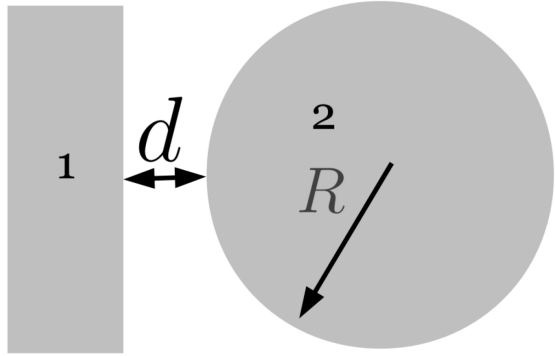}
	\caption{\label{Fig:platesphere}Sketch of the plate-sphere setup considered here (not to scale). $R$ denotes the radius of the sphere and $d$ the distance of closest approach between the sphere and the plate. In this case $R\gg{}d$ so that the PFA can be used to neglect the curvature of the sphere.}
\end{figure}
Since this paper focuses on the calculation of the Casimir force at separations below 100 nm, where surface roughness and optical properties play important roles, its temperature dependence can be ignored.\cite{Milton2004} The starting point of our calculations is the macroscopic Casimir-Lifshitz energy per unit area between parallel dielectric plates separated by a vacuum gap of width $d$ in the low temperature limit where $k_bT\ll\hbar{}c/2d$:\cite{Lifshitz61}
\begin{equation}\label{Lifshitz}
E(d) = -\frac{\hbar}{16\pi^2d^2}\sum\limits_{\mu = s, p}\int\limits_{0}^{\infty}\mathrm{d}\zeta\int\limits_{\zeta/\zeta_c}^{\infty}x^2\mathrm{d}x\ln(1-R_{\mu}e^{-x}),
\end{equation}
where $x=2k_0d$, $k_0=\sqrt{\zeta^2/c^2+q^2}$, $q$ denotes the radial wavenumber, $\zeta$ the imaginary part of the frequency, and $\zeta_c\equiv{}c/2d$ the characteristic frequency. Finally, $R_\mu=r_{1\mu}r_{2\mu}$ denotes the product of the Fresnel reflection coefficients for plate 1 ($r_{1\mu}$) and plate 2 $(r_{2\mu})$, given by:
\begin{eqnarray}\label{Fresnel}
r_{js}=\frac{k_0-k_j}{k_0+k_j}\\
r_{jp}=\frac{\varepsilon_j(i\zeta)k_0-k_j}{\varepsilon_j(i\zeta)k_0+k_j},\nonumber
\end{eqnarray}
where the subscript $\mu=s,p$ denotes the polarization and the index $j=1,2$ labels the bodies. The permittivities at imaginary frequencies can be obtained from the ones at real frequencies via the Kramers-Kronig relations:

\begin{figure}[!ptb]
	\centering
		\includegraphics[width=0.45\textwidth]{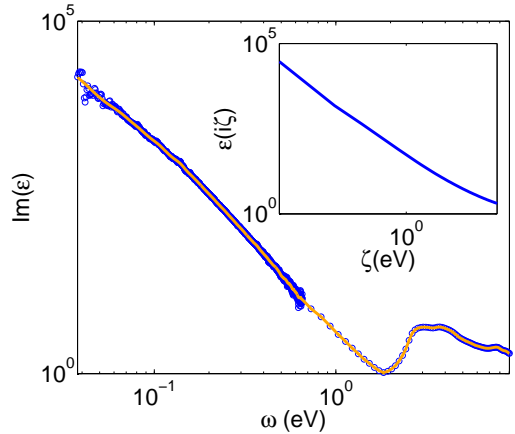}		
	\caption{\label{Fig:ellipsometry}(color online) Ellipsometry data of a gold sample. The main panel shows the imaginary part of the dielectric function. The (blue) circles represent ellipsometry measurements, and the continuous (orange) line is a smoothed fit, which was used in the calculations. The inset is a plot of the permittivity at imaginary frequencies $\varepsilon(i\zeta)$ as obtained from the Kramers-Kronig relations. The latter function enters the Lifshitz formula for the Casimir force calculations between smooth surfaces.}
\end{figure}

%
\begin{equation}\label{KK}
\varepsilon(i\zeta)=1+\frac{2}{\pi}\int\limits_{0}^{\infty}\mathrm{d}\omega\frac{\omega\mathrm{Im}(\varepsilon(\omega))}{\omega^2+\zeta^2}.
\end{equation}
Calculation of the Casimir force requires knowledge of the imaginary parts of the permittivities in a broad frequency range. For this purpose we used ellipsometry data for the frequency dependent permittivities of Au surfaces in the range of 0.038 to 9.85 eV (see Fig. \ref{Fig:ellipsometry}). We have extrapolated to frequencies below 0.038 eV with the Drude model:
\begin{equation}\label{Drude}
\varepsilon(\omega)=1-\frac{\omega_p^2}{\omega(\omega+i\omega_\tau)},
\end{equation}
where the values of the plasma frequency $\omega_p$ and the relaxation parameter $\omega_\tau$ were:\cite{PeterOptical} $\omega_p=7.8$ eV, $\omega_\tau=49$ meV.
\par Finally, in order to compare to experimental results, we give the corresponding expression for the force. Experiments are commonly performed in a sphere-plate configuration to avoid problems with the alignment between parallel plates (Fig. \ref{Fig:platesphere}). If the radius of the sphere $R$ is much larger than the separation $d$, the PFA can be used to neglect the effect of the sphere's curvature on the Casimir force via
\begin{equation}\label{PFAcurv}
F(d)=2\pi{}RE(d)\qquad{}R\gg{}d,
\end{equation}
where $E(d)$ is given by Eq. \eqref{Lifshitz}. In a plate-plate configuration this approximation is not necessary and  the Lifshitz formula \cite{Lifshitz61} provides an explicit expression for the force: $F(d) = -E'(d)A$, where $A$ is the surface of each plate.
\subsection{\label{statistics}Extreme value statistics and contact distance}
\begin{figure*}[!ptb]
\centering
	\subfigure{\includegraphics[width=0.2\textwidth]{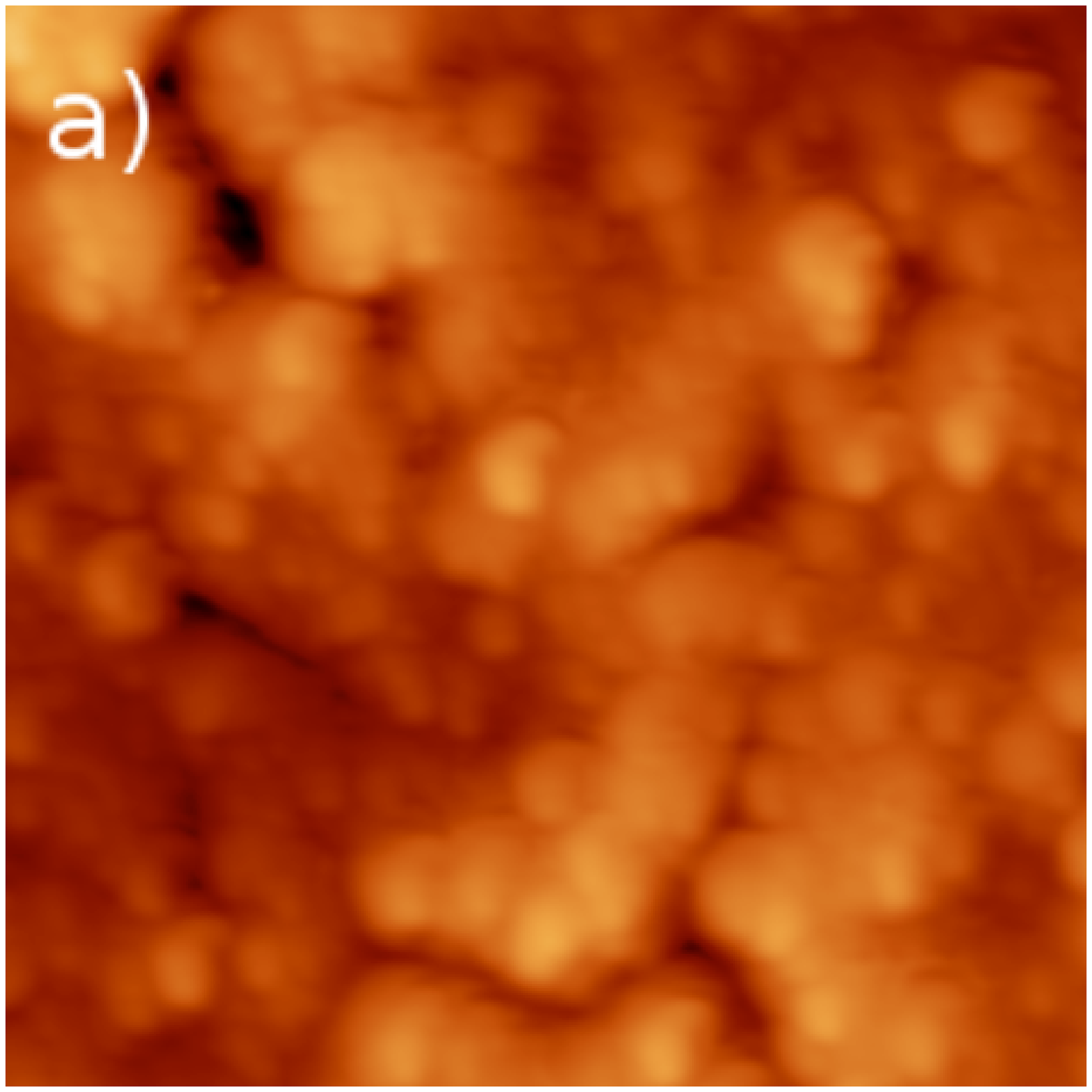}\label{Fig:AFMscansa}}\quad
		\subfigure{\includegraphics[width=0.2\textwidth]{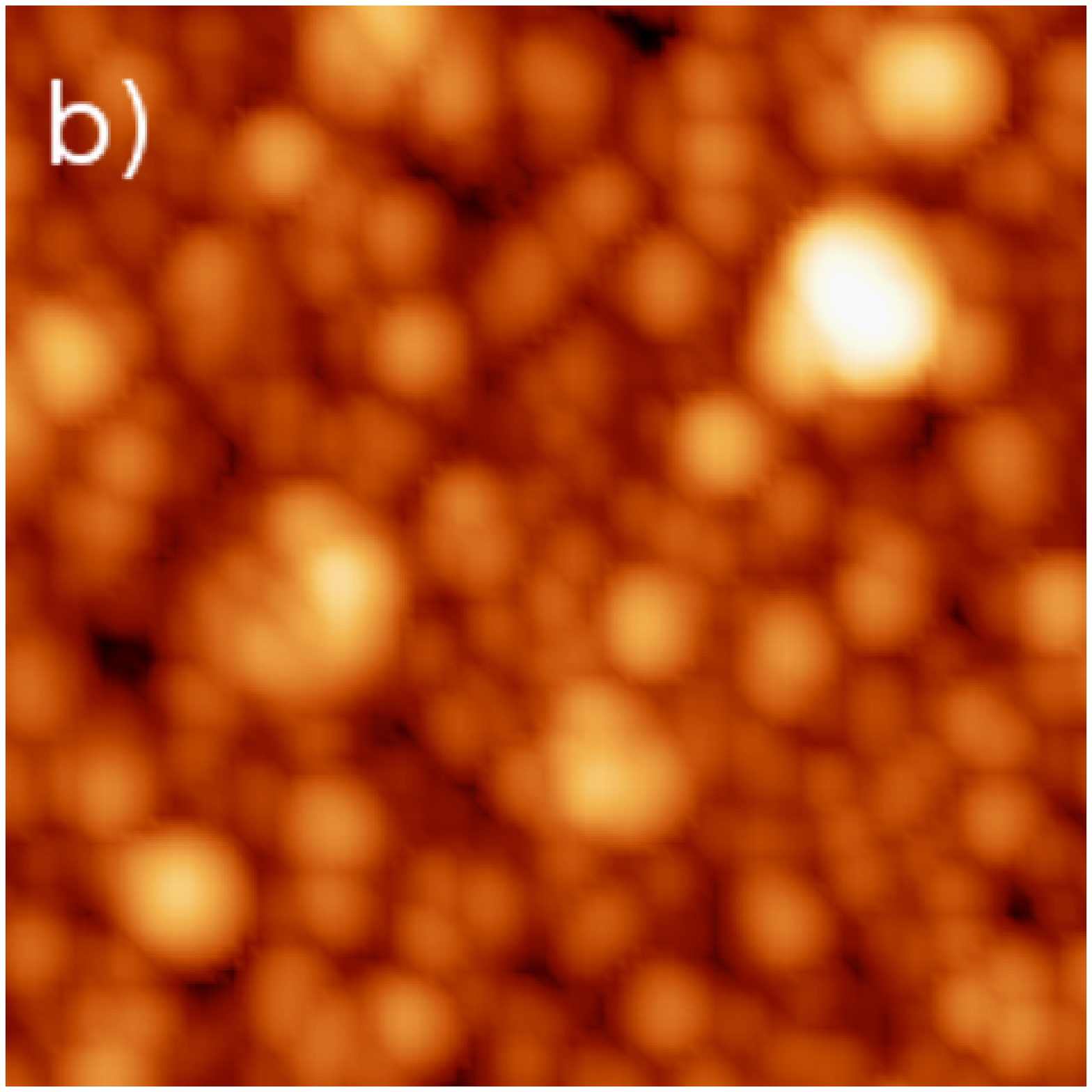}\label{Fig:AFMscansb}}\quad
		\subfigure{\includegraphics[width=0.2\textwidth]{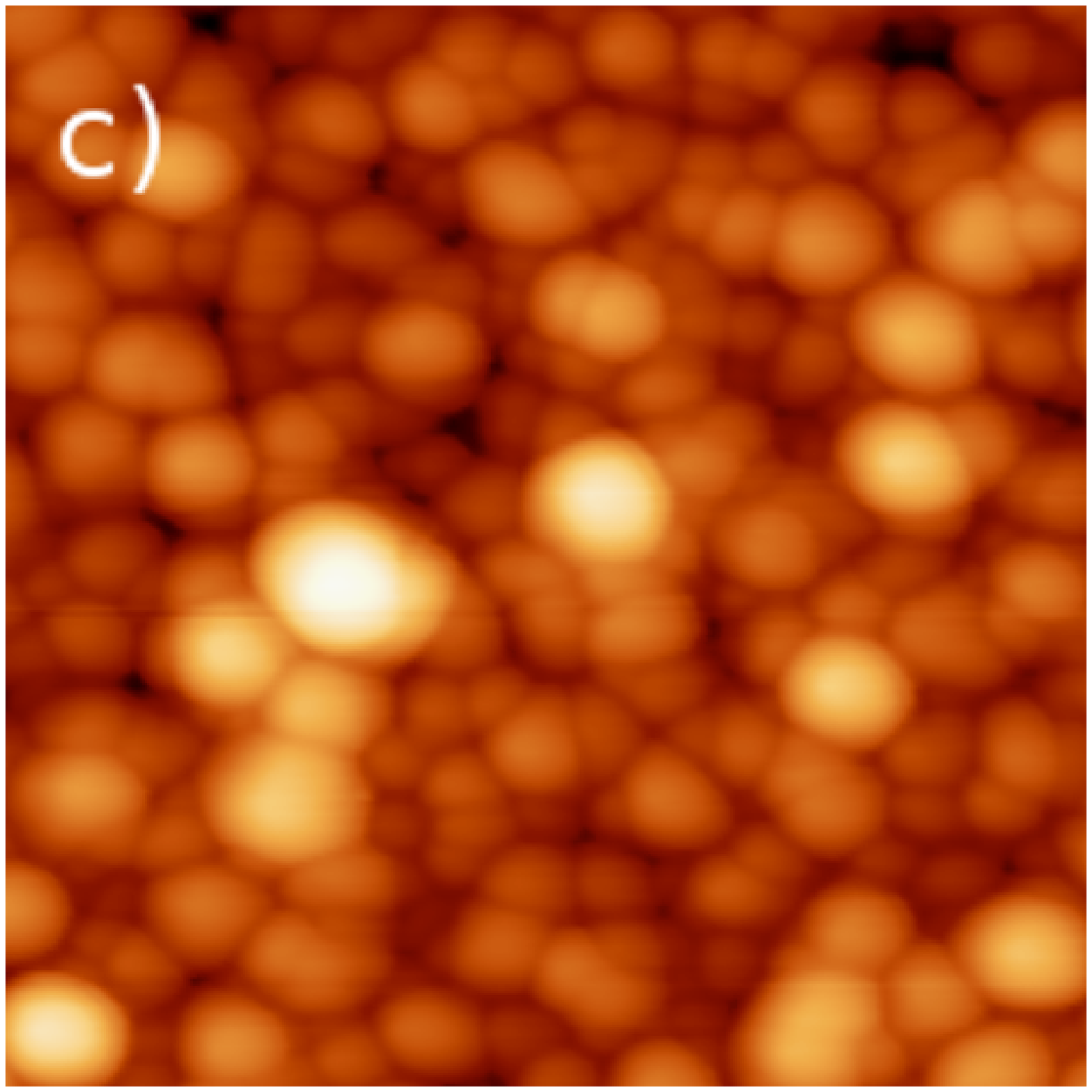}\label{Fig:AFMscansc}}\quad
		\subfigure{\includegraphics[width=0.2\textwidth]{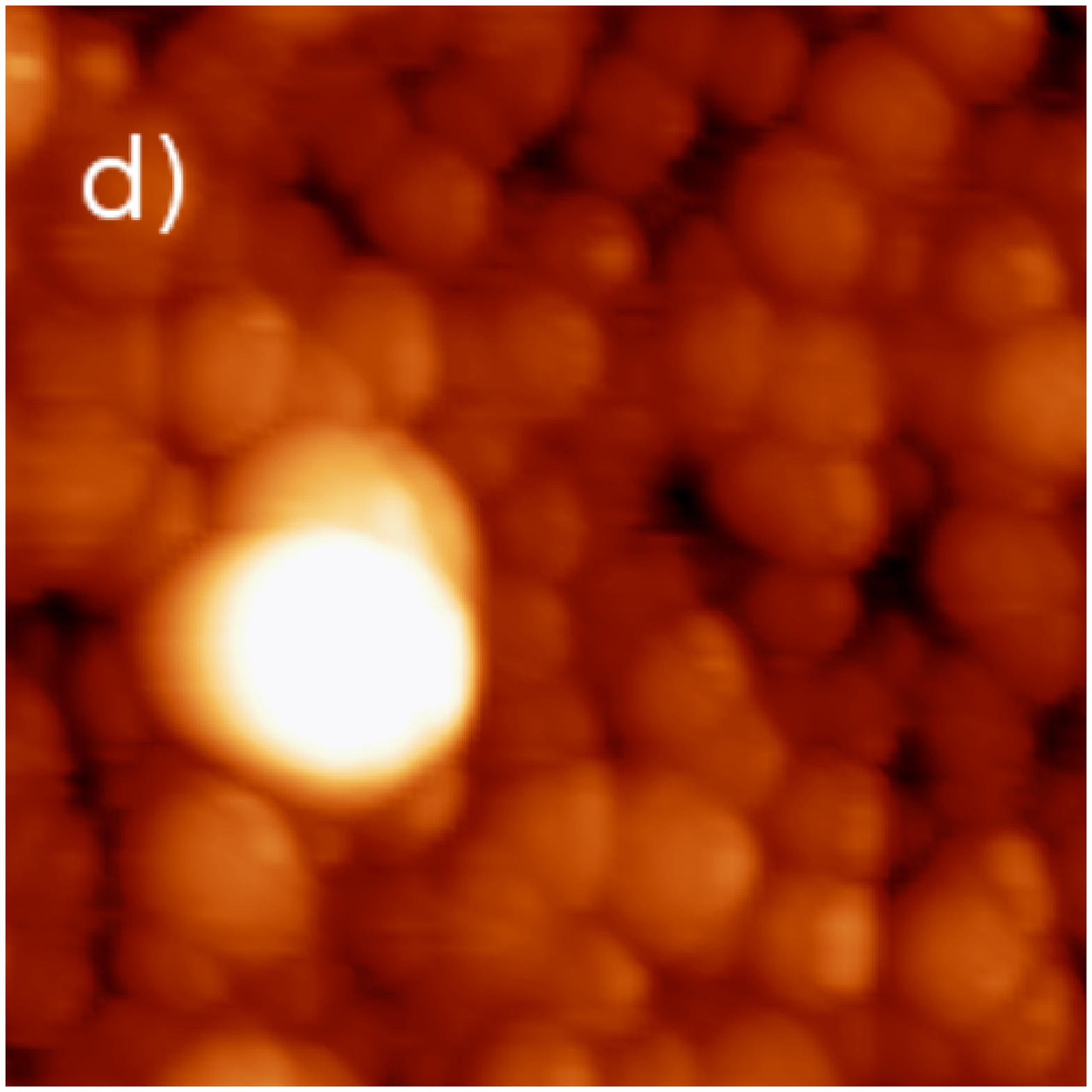}\label{Fig:AFMscansd}}\quad
		\subfigure{\includegraphics[scale=0.25]{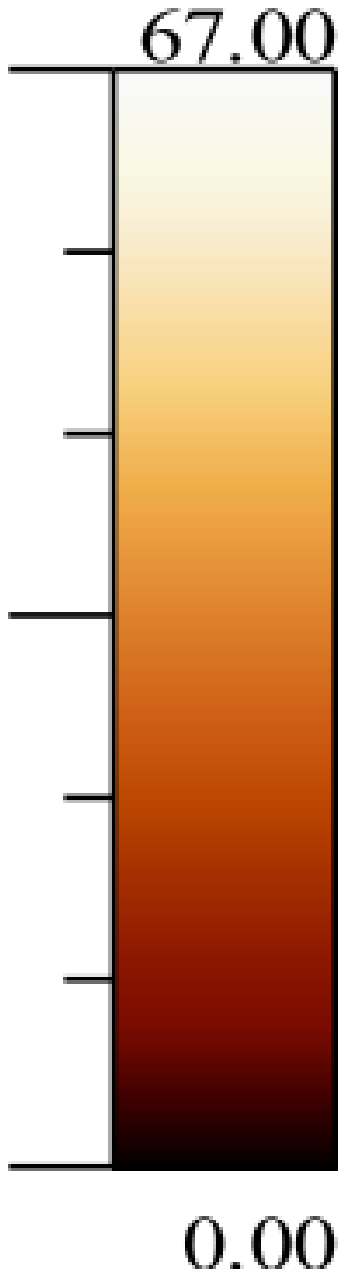}}		
			\protect\caption{\label{Fig:AFMscans}(color online) AFM scans of the samples used in the calculations. An area of 1$\mu$m$\times1\mu$m is shown in each picture. The parameters for the actual calculations were: For the sphere (shown in \subref{Fig:AFMscansa}) an area of 8$\mu$m$\times$8$\mu$m was scanned at a resolution of 2048$\times$2048 pixels. For the 1600 nm sample (shown in \subref{Fig:AFMscansb}) this area was 40$\mu$m$\times$40$\mu$m at a 2944$\times$2944 resolution. The scan size for the 1200 nm sample (Fig. \subref{Fig:AFMscansc}) was 10$\mu$m$\times$10$\mu$m at a 2048$\times$2048 resolution. The area and resolution used for the 800 nm sample (Fig. \subref{Fig:AFMscansd}) are respectively 40$\times$40$\mu$m$^2$ and 4096$\times$4096 pixels. The color bar indicates the vertical scale in nm.{}}
	
\end{figure*}

\begin{figure*}[!ptb]
	\centering
		\subfigure{\includegraphics[width=0.49\textwidth]{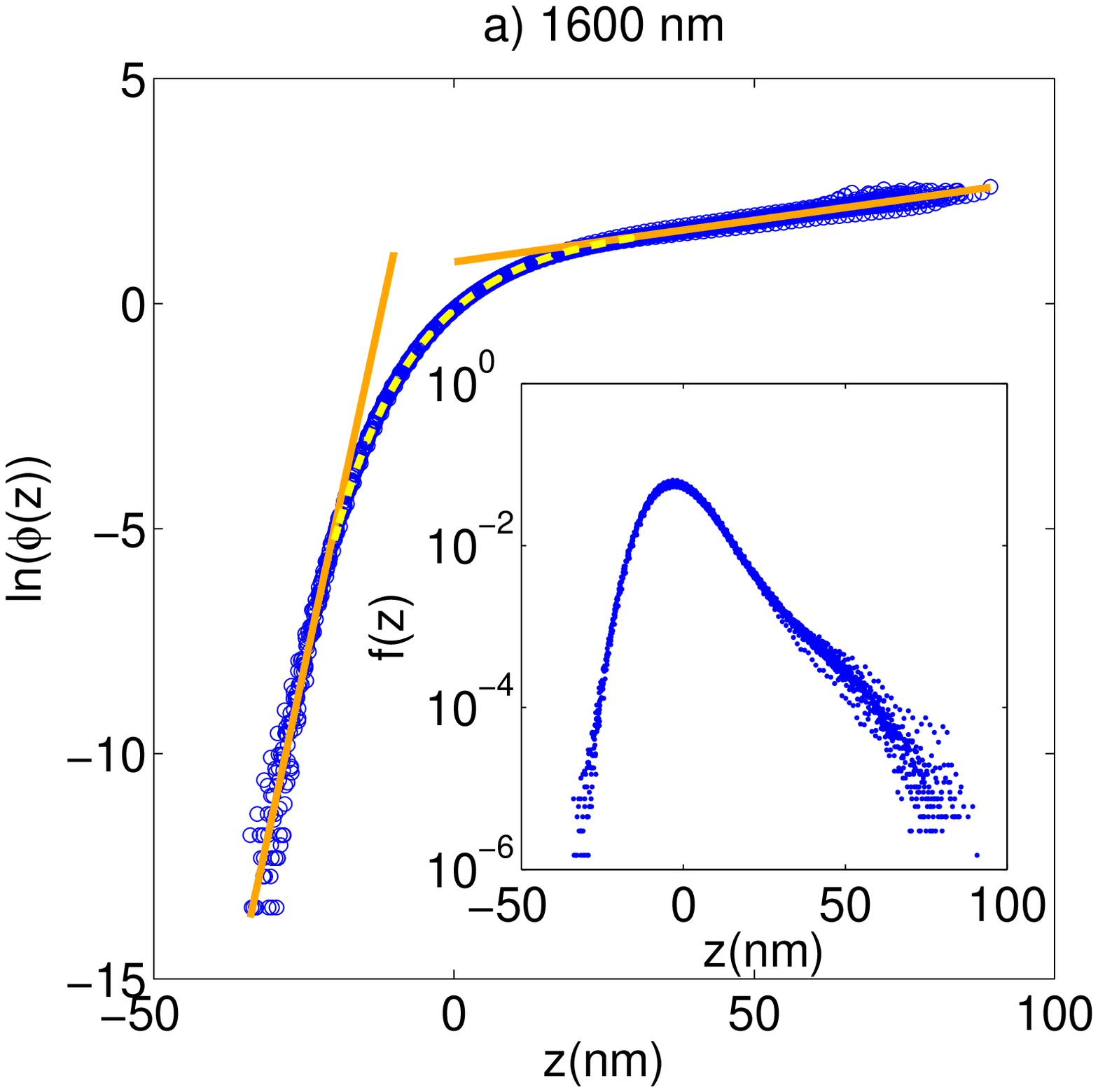}\label{Fig:distributiona}}
		\subfigure{\includegraphics[width=0.49\textwidth]{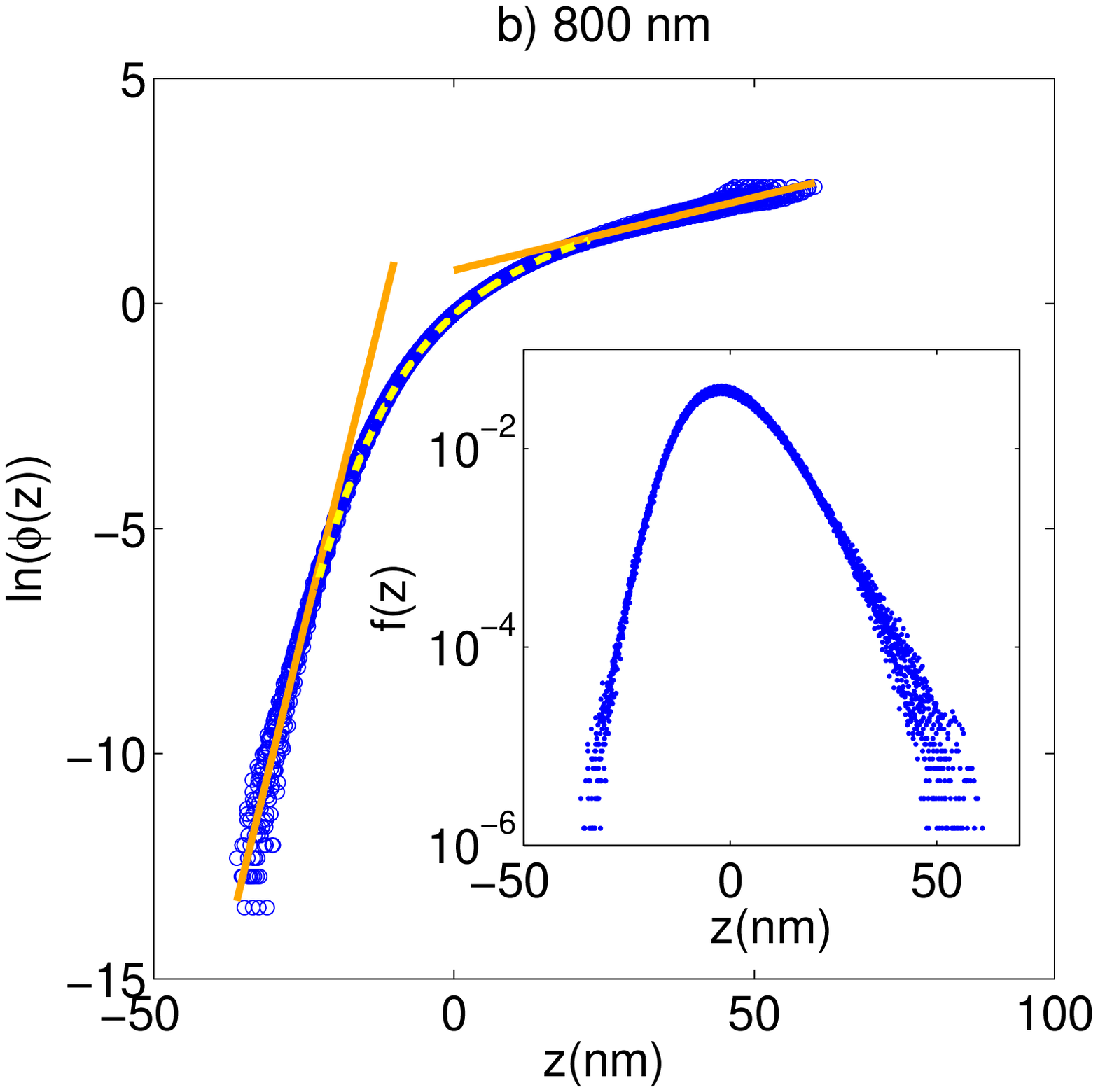}\label{Fig:distributionb}}
	\caption{(color online) Statistics of the topography of the surfaces. The logarithm of the ``phase'' is plotted as a function of the height with respect to the mean plane $z=0$. The open circles represent the data from AFM topography scans: \subref{Fig:distributiona} is for the 1600 thick film and \subref{Fig:distributionb} the 800 nm thick one. The solid (orange) lines represent linear fits for $|z|\gg1$. This implies that the probability to find a large feature behaves as a Gumbel distribution. For intermediate values of $z$ the data are fit with polynomials, indicated by the dashed (yellow) curves. The inset is a semilogarithmic plot of the probability density function $f(z)$. It shows significant deviation from  a normal distribution for the 1600 nm sample. The distribution of the 800 nm sample deviates less from a normal distribution, but it is still clearly asymmetric.}
	\label{Fig:distribution}
\end{figure*}

Assessing the influence of random surface roughness on the Casimir force requires knowledge of the proper probability distributions of the height fluctuations of the surfaces.  These were obtained from AFM scans of each film with lateral resolutions varying from 4 to 10 nm, for areas up to $40\times40\mu{}m^2$. (See Fig. \ref{Fig:AFMscans} for detailed parameters.)
This information enables us to perform a detailed roughness analysis of the samples. By counting the number of features smaller than some value $z$ and normalizing this number, the cumulative probability $P(z)$ to find a feature smaller than $z$ is obtained. It turns out that this probability approaches $1$ very fast at $z\rightarrow\infty$ and 0 $z\rightarrow-\infty$. This is why, for a proper analysis of the AFM data, it is convenient to write $P(z)$ as:
\begin{equation}\label{dist_P}
P(z)=1-e^{-\phi(z)},
\end{equation}
where the ``phase'' $\phi(z)$ is a positive, monotonically increasing function of $z$, defined as
\begin{equation}\label{dist_phi}
\phi(z)\equiv-\ln(1-P(z)).
\end{equation}
The derivative of $P(z)$,
\begin{equation}
f(z)\equiv{}P'(z)=(1-P(z))\phi'(z),
\label{eq:fzdef}
\end{equation}
is the probability density function. It was established \cite{d0paper} that $\phi(z)$ could not be fitted to any known distribution for all $z$ and that for large $|z|$ a generalized extreme value distribution is needed. Figure \ref{Fig:distribution} shows the natural logarithm of  the phase $\phi(z)$, collected from the AFM images. It is clear that this function behaves linearly in the asymptotic regimes:
\begin{equation}\label{pos_asym}
\ln\phi(z)=A_{+}z+B_{+},
\end{equation}
for large positive $z$ and similarly,
\begin{equation}\label{neg_asym}
\ln\phi(z)=A_{-}z+B_{-}
\end{equation}
for large negative $z$. The values of the coefficients $A_{\pm}$ and $B_{\pm}$ are listed in table \ref{tab:parameters}.
This linear behavior in the asymptotic regimes implies that the probability to find a feature larger than $z$ behaves as a `double exponential':
\begin{equation}\label{Gumbel}
1-P(z)\sim{\exp(-\exp(\tfrac{z-\mu}{\beta}})), 
\end{equation}
where $\beta$ is the scale parameter and $\mu$ is the location parameter. This type of behavior is a characteristic of Gumbel distributions, which is an example of extreme value statistics. \cite{Gumbelbook} We will see that this strong dependence will have a considerable impact on the roughness correction to the Casimir force. In this paper we have analyzed only gold samples and we cannot draw conclusions for other materials. However, the generality of the Gumbel distribution allows us to hope that similar behavior can be found in the roughness statistics of other materials.
\par
\begin{figure}[!ptb]
	\centering
		\includegraphics[width=0.45\textwidth]{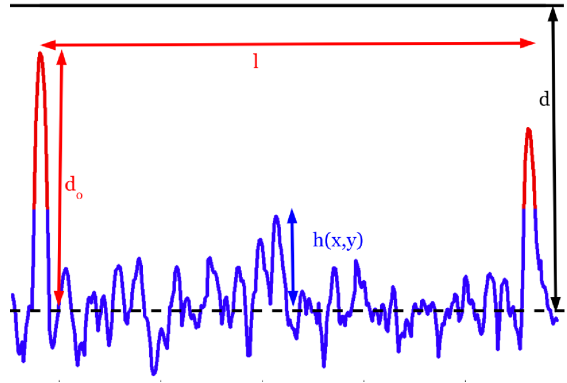}
	\caption{(color online) Schematic of a rough surface to clarify the meaning of the parameters $l$ and $d_0$. Similar to $l$, $l'$ represents the distance between the deep troughs, and $d_0'$ represents the depth of the deepest pit. $h(x,y)=h_1(x,y)+h_2(x,y)$ is the combined surface profile, so that effectively only one rough surface is considered.}
	\label{Fig:setup}
\end{figure}

  The distance upon contact for gold films was discussed in detail in Ref. \onlinecite{d0paper}. The thicknesses of the investigated gold films are  associated with different rms roughnesses due to the kinetic roughening process. We denote the height fluctuations from the mean surface level by $h_j(x, y)$  for each body ($j=1,2$). The local separation distance is $d_{\mathrm{local}}(x,y)=d-h_1(,x,y)-h_2(x,y)$. The averages over a large surface of each profile is zero by definition: $<h_j(x,y))> = 0$. Another assumption is that the surfaces are statistically independent, i.e. the surface heights are uncorrelated:
\begin{equation}\label{surfacesindep}
\left\langle h_1(x_1,y_1)h_2(x_2,y_2)\right\rangle = 0
\end{equation}
which is a condition for a perturbative treatment.\cite{Genet2003, Lambrecht2005} Consequently, the profiles can be combined so that effectively one rough body with topography $h(x,y)= h_1(x,y)+h_2(x,y)$ is considered, interacting with a flat  surface (see Fig. \ref{Fig:setup}). In the plate-sphere configuration, the contact distance is defined as\cite{d0paper} the maximum average separation $d$ for which the local distance becomes zero, so that
\begin{equation}\label{d0def}
d_0(L)\equiv\max_{x,y}\left[h(x,y)-(x^2+y^2)/2R\right],
\end{equation}
where $L$ denotes the size of the effective interaction area. The contact distance is the local maximum within the horizontal scale $L$. In a plate-plate geometry $R\rightarrow\infty$ and $d_0=\max{}\left[h(x,y)\right]$. Throughout this paper it is assumed that the sphere is fixed laterally with respect to the plate and that it does not rotate during force measurements (in reality it is rigidly attached to a cantilever). In other words, we distinguish the \emph{experimental} uncertainty in $d_0$ from its \emph{statistical} uncertainty. Indeed, if the sphere is allowed to move laterally, the uncertainty in the value of $d_0$, and therefore in the Casimir force will be considerably larger.\cite{d0paper}
\par In this paper it is assumed that the size of the effective interaction area between the sphere and the plate $L$ is much larger than the correlation length: $L\gg\xi$. This ensures that one interaction area contains many independent realizations of a rough surface and hence spatial averages are equivalent to statistical averages. Our approach requires a large size of the plate also on the scale of the separation $d$: the condition $L\gg{}d$ ensures that edge effects can be ignored. These conditions for $L$ are realistic: $L$ is in the order of a few microns, while $d$ and $\xi$ are a few tens of nm. (See table \ref{tab:parameters}.)

%
\begin{table*}
\caption{\label{tab:parameters}Values of the relevant parameters for the investigated gold samples. The parameters $A_{\pm}$ and $B_{\pm}$ are defined by Eqs. \eqref{pos_asym} and \eqref{neg_asym}. The subscripts $+$ and $-$ refer to the positive and negative asymptote, respectively. The other parameters are: the rms roughness $w$ {of the sample}, the correlation length $\xi$, the size of the effective interaction area $L$, the contact distance $d_0$, and the average distance between the high peaks $l$.}
\begin{ruledtabular}
\begin{tabular}{cccccccccc}
 &\multicolumn{2}{c}{}&\multicolumn{2}{c}{}\\
Thickness (nm)&$A_+$ (nm$^{-1}$)&$B_+$&$A_{-}$(nm$^{-1}$)
&$B_{-}$ & $w$ (nm) & $\xi$ (nm) & $L$ (nm)& $d_0$ (nm) & $l$ (nm)\\ \hline
800&0.0333&0.704 &{0.542}&{6.34}& 7.5 & 30.6 & 1560 & 34.5 $\pm$ 1.7 & 238\\
1200&0.0188&0.888&0.648&8.34 & 9.0 & 38 & 1980&41.0 $\pm$ 1.7 & 256\\
 1600&0.0192&0.885&{0.62} & {7.32} & 10.1 & 42.0 & 2100 & 50.8 $\pm$ 1.3 & 380\\
\end{tabular}
\end{ruledtabular}
\end{table*}
%

\section{Model outline}\label{model}

A rough surface can be regarded as a large number of asperities of
different heights typically $\sim w$  and lateral sizes $\xi$ with a
few occasional high peaks.{Here $w$ is the total root mean square roughness defined as $\sqrt{w_{\mathrm{sphere}}^2+w_{\mathrm{sample}}^2}$}. The asperities with the heights $\sim w$
can be well described by a normal distribution. This is clear
from the insets in Fig. \ref{Fig:distribution}: the function $\ln
f(z)$ can be approximated by a parabola nearby its maximum. However,
the tails of the distribution, which correspond to high peaks or
deep troughs, cannot be described by the normal law. Let us define
the parameter $d_1$ in such a way that asperities with normal heights
are smaller than $d_1$, $h<d_1$, but the high peaks are larger than
$d_1$, $h>d_1$. The value of $d_1$ belongs to the interval
$w<d_1<d_0$. Its precise value is somewhat up to convention but it can be
chosen around $d_1\sim 3w$. An additional condition on $d_1$ will be
discussed later in this section.

The number of high peaks with the lateral size $\xi$ and the height
$h>d_1$ on the area $L^2$ can be expressed via the "phase" $\phi(z)$
determined from the roughness topography as
\begin{equation}\label{model7}
N=\frac{L^2}{\xi^2}e^{-\phi(d_1)}.
\end{equation}
The average distance between these peaks (Fig. \ref{Fig:setup}) is
\begin{equation}\label{model8}
l=\frac{L}{\sqrt{N}}=\xi{}e^{\phi(d_1)/2}.
\end{equation}
Similarly we say that the deep troughs are those having depths larger
than $d_1'$, $h<-d_1'$. The number of these troughs on the area $L^2$
is
\begin{equation}\label{model9}
N'=\frac{L^2}{\xi^2}\phi(-d_1')
\end{equation}
and the average distance between them is
\begin{equation}\label{model10}
l'=\frac{L}{\sqrt{N'}}=\frac{\xi}{\sqrt{\phi(-d_1)}}.
\end{equation}

Consider first the roughness contribution to the Casimir force in the
case of large correlation length $\xi\gg d$. In this limit PFA is a
good approximation \cite{Genet2003} in the sense that each asperity
can be taken into account independently (additively). Then we can
calculate the Casimir force $F_{Cas}(d)$ via the standard definition
of the statistically averaged function
\begin{equation}\label{model1}
F_{Cas}(d)=\int\limits_{-\infty}^{\infty}{\mathrm{d}zf(z)F(d-z)}.
\end{equation}
Here we defined $f(z)=0$ outside the interval $-d_0'< z < d_0$.
Writing the force as an integral over the entire real axis is useful
to obtain a result in terms of statistical parameters, such as $w$.
If additionally the distance between bodies is large in comparison
with the roughness, $d\gg w$, we can expand the force between flat
plates around $z=0$ as $F(d-z)=F(d)-F'(d)z+F''(d)z^2/2+\ldots$ and
find the roughness correction:
\begin{eqnarray}\label{model2}
F_{Cas}(d)=F(d)+\frac{F''(d)}{2!}w^2+\ldots,\ \ \ w\ll d.
\end{eqnarray}
which is the second term in (\ref{model2}). The error due to omitted
terms can be estimated via the approximate power law dependence of
$F(d)$ on $d$ in Eq. \eqref{PFAcurv}. \cite{PeterIjmpB2010}

Let us separate three different integration intervals in Eq.
\eqref{model1}:
\begin{equation}\label{model4}
    F_{Cas}(d) = \int\limits_{-\infty}^{-d_1'}{\mathrm{d}z f(z)F(d-z)} +
    \int\limits_{-d_1'}^{d_1} \ldots +
    \int\limits_{d_1}^{\infty} \ldots,
\end{equation}
where $\ldots$ stands for $\mathrm{d}z f(z)F(d-z)$. The first term
here represents  the contribution of deep troughs, the second one is
responsible for the contribution of normal peaks, and the third term
is the contribution of high peaks. An important observation of this
work is that the contribution of normal peaks with the height $\sim
w$ can be taken into account perturbatively even if the bodies are
already in contact. It follows from the fact \cite{d0paper} that upon
the contact the bodies are still separated by the distance $d_0$,
which increases with the area of nominal contact and is in the range
$3w\leq d_0 \leq 5w$. In this case the Taylor expansion for $F(d-z)$ in the
second term is justified \cite{BroerEPL2011}.

Now let us relax the condition $\xi\gg d$. In this case we cannot
consider different asperities as independent. The method to calculate
the roughness correction beyond the PFA was proposed in the series of
papers \cite{Genet2003, Lambrecht2005}. In this approach the
roughness is treated perturbatively. We can apply this method only to
the second term in (\ref{model4})
\begin{equation}\label{2dterm}
    \int\limits_{-d_1'}^{d_1}{\mathrm{d}z f(z)F(d-z)}=
    \int\limits_{-\infty}^{\infty}\ldots+
    \int\limits_{-\infty}^{-d_1'}\ldots+
    \int\limits_{d_1}^{\infty}\ldots,
\end{equation}
where we have to understand the function $F(d-z)$ as the Taylor
expansion. According to \cite{Genet2003, Lambrecht2005} the first
term on the right has to be generalized in the following way
\begin{eqnarray}\label{model6}
    \nonumber F_{PT}(d) & \equiv & \int\limits_{-\infty}^{\infty}
    {\mathrm{d}z f(z)\left[F(d)-F'(d)z+\frac{F''(d)}{2!}z^2\right]}\\
    &&\rightarrow F(d)+\frac{F''(d)}{2!}\int{\frac{\mathrm{d}^2k}{(2\pi)^2}\;\rho(kd)\sigma(k)},
\end{eqnarray}
where $\sigma(k)=\left\langle{}h(k)h(0)\right\rangle$ is the
correlator of the surface profile in $k$-space. The sensitivity
function, $\rho(kd)$, is defined as the ratio between the response
functions at arbitrary  and at zero wavenumber: $\rho(k)\equiv
G(k)/G(0)$. It measures the deviation from the PFA. The proximity
force approximation is restored when small wavenumbers $kd\ll 1$ are
important (large $\xi$). In this case the sensitivity function is
$\rho(kd)\rightarrow 1$ and we reproduce Eq. (\ref{model2}). The
expression for the function $\rho(kd)$ is given in
\cite{Genet2003,Lambrecht2005}. It has to be noted that $\rho\geq1$,
thus, the PFA underestimates the Casimir force.

When the condition $\xi\gg d$ is broken we are able to calculate the
second term in (\ref{model4}) by using the perturbation theory but we
definitely cannot use the perturbation theory for the third term.
This is because at $z=d_0$ the integrand diverges (for $z>d_0$ we
defined $f(z)=0$). This is a physical divergence appearing due to the
contact between the highest asperity and the opposite body. However,
it can be noted that the high peaks accounted  for by the third term in
(\ref{model4}) are rare and the average distance between them
(\ref{model8}) is large. If this distance is large in comparison with
the separation between bodies, $l\gg d$, we can calculate the
contribution of each peak independently of each other (additively).
We can always choose $d_1$ to fulfill the condition $l\gg d$ but in
reality $d_1=3w$ is an appropriate value in all respects. As one can
see from Table \ref{tab:parameters} for all the investigated films
the values of $l$ are sufficiently large and the values of $d_1$ are
always smaller than $d_0$. It is also important that our results are
not sensitive to the precise value of $d_1$ as long as $d_1$ is
around $3w$. This is clear from the insets in Fig.
\ref{Fig:distribution}: there is no sharp point in the function
$f(z)$ where the normal distribution becomes inapplicable.

The precise value $d_1'$ for the deep troughs is not important at
all. Any value in the interval $w<d_1'<d_0'$ is equally good. This is
mainly because the contribution of the deep troughs is small and
never dominates but also due to the fact that $\ln\phi(z)$ decreases more
sharply at large negative $z$ than  it increases at large positive $z$.

The discussion above shows that the high peaks and deep troughs can
be calculated additively even in the case when applicability of the
PFA is unjustified. In this case instead of (\ref{model4}) we can
write
\begin{eqnarray}\label{model11}
\nonumber &&F_{Cas}(d)=F_{PT}(d)+\\
\nonumber && \int\limits_{d_1}^{\infty}{\mathrm{d}z f(z)\Big[F(d-z)-F(d)+F'(d)z-\frac{F''(d)}{2!}z^2\Big]}+\\
\nonumber && \int\limits_{-\infty}^{-d_1'}{\mathrm{d}z f(z)\Big[F(d-z)-F(d)+F'(d)z-\frac{F''(d)}{2!}z^2\Big]}.
\end{eqnarray}
where the remnants from $F_{PT}(d)$ in Eq. (\ref{2dterm}) are
included in the terms responsible for high peaks and deep troughs.
The final expression for the force is split into three terms:
\begin{eqnarray}\label{modelfinal}
F_{Cas}(d)=F_{PT}(d)+F_{PFA}(d)+F_{PFA}'(d).
\end{eqnarray}
The first term,
\begin{equation}\label{FPT}
F_{PT}(d)=F(d)+\frac{F''(d)}{2!}\int{\frac{\mathrm{d}^2k}{(2\pi)^2}\rho(kd)\sigma(k)},
\end{equation}
does not rely on the PFA but is instead based on the perturbation theory
\cite{Genet2003, Lambrecht2005} as indicated by the index $PT$. It
represents the contribution of asperities with typical heights $\sim
w$. The second term,
\begin{eqnarray}\label{FPFA}
F_{PFA}(d)=\int\limits_{d_1}^{d_0}\mathrm{d}z f(z)\Big[F(d-z)-\nonumber\\
F(d)+F'(d)z-\frac{F''(d)}{2!}z^2\Big],
\end{eqnarray}
is the contribution of high peaks. In this term the perturbation
theory cannot be used to calculate $F(d-z)$ because $d$ and $z$ are
comparable. $F_{PFA}(d)$ diverges at $d=d_0$. As was already
mentioned this is because the local separation distance becomes zero
at $d=d_0$. In this way the model accounts for the case of contact
between the bodies. This will turn out to be an important aspect of
our approach. The condition $L\gg \xi$ ensures that the interaction
area contains enough realizations of a rough surface to approximate
an ensemble. \emph{Since the high peaks are statistically rare events
they should be far apart, $l\gg\xi$, so that they can be calculated
independently of each other.}  Previously we assumed
\cite{BroerEPL2011} that the high peaks have flat tips, so that one
can use the PFA to calculate the interaction between an individual
peak and a flat surface. This approximation is reasonable (see
\cite{BroerEPL2011}) but it is not necessary and we relax it in
section \ref{curvature}.

Finally, the term
\begin{eqnarray}\label{FPFA'}
F_{PFA}'(d)=\int\limits_{-d_0'}^{-d_1'}\mathrm{d}zf(z)\Big[F(d-z)-F(d)+\nonumber\\ F'(d)z-\frac{F''(d)}{2!}z^2\Big]
\end{eqnarray}
represents the contribution of the deep troughs. By the same token,
the distance between them is large, so that their contributions are
also independent of each other. These troughs do not dominate the
force, because they correspond to negative $z$, where the leading
term $F(d-z)$ is much smaller than the other contributions.

\section{\label{curvature}The influence of the shape of the peaks}
%
%
\begin{figure*}[!pt]		\subfigure{\includegraphics[width=0.45\textwidth]{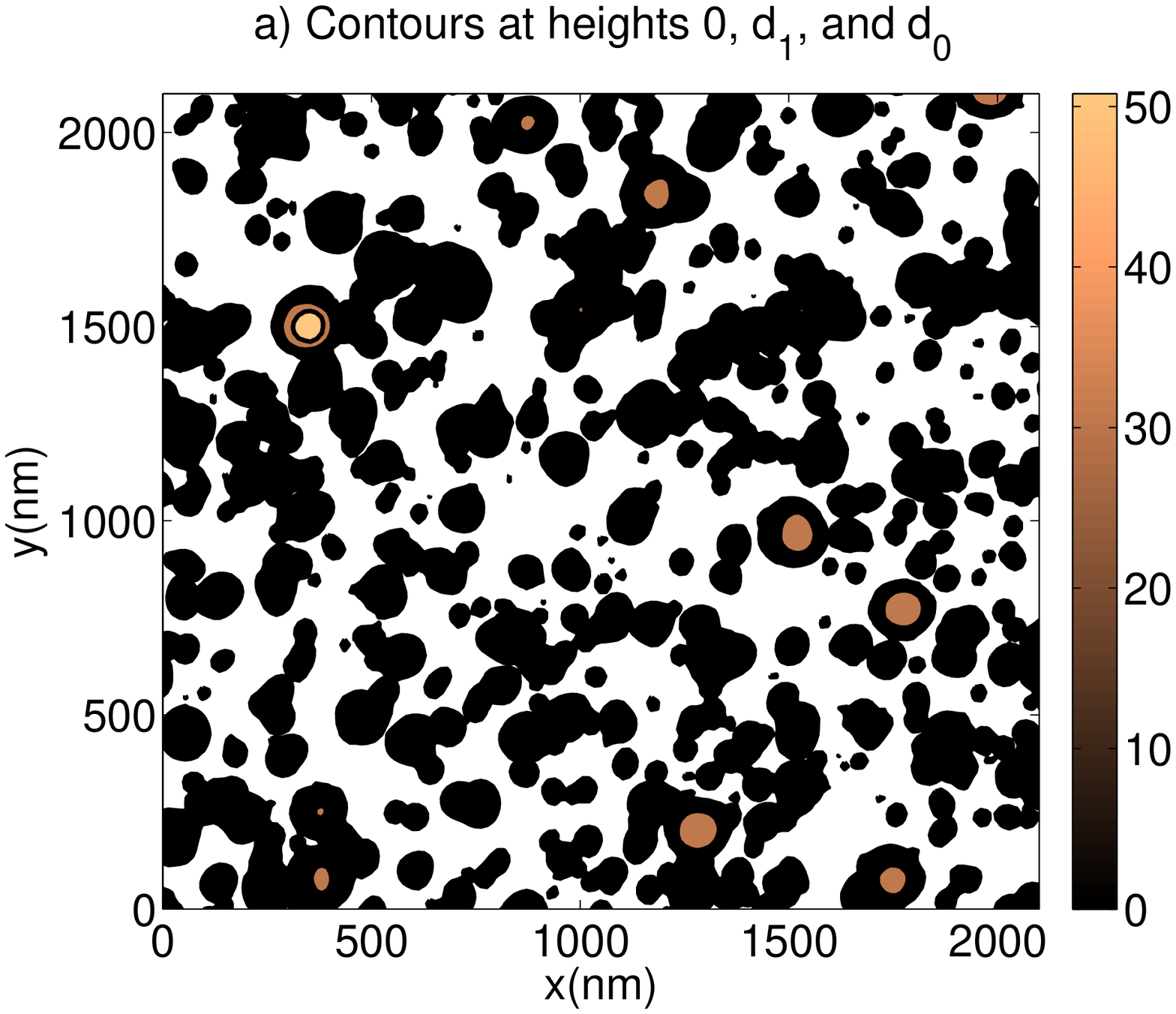}\label{fig:contour1600a}}		\subfigure{\includegraphics[width=0.45\textwidth]{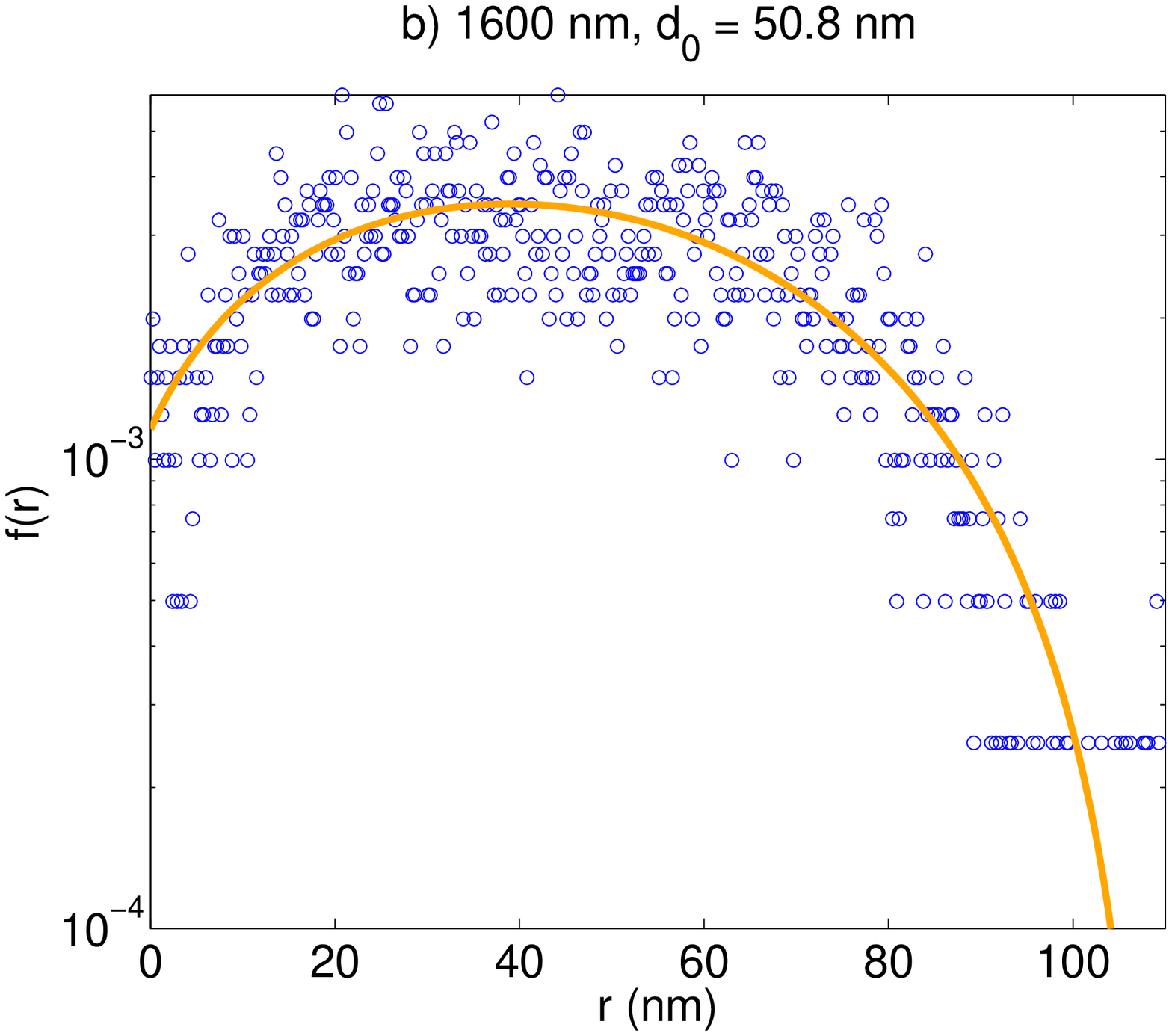}\label{fig:contour1600b}}
	\caption{(color online) \subref{fig:contour1600a} Contour plot of subsurface of size $L^2$ for the 1600 nm sample, extracted from AFM data. The colorbar indicates the vertical scale in nm. At height $z=d_1=$ 30.3 nm, the polygons can be considered circular by approximation. \subref{fig:contour1600b} Semilogarithmic plot of probability density function $f(r)$ of the radii of the peaks. The open circles represent data from the AFM topography scan, the (orange) line is a polynomial fit. This information is used to estimate the range of the horizontal sizes of the peaks. 
	}
	\label{fig:contour1600}
\end{figure*}
{In order determine the effect of the shape of the peaks one must first establish what geometry approximates the shape of the real peaks best. We note that the rough surface in the schematic of Fig. \ref{Fig:setup} is a cross section of a real rough gold surface  based on an AFM scan of the 1600 nm sample. At present it does not seem feasible to determine the shape of the peaks directly from this image since the size of the tip of the AFM cantilever beam is comparable to the size of the tips of the peaks.}
\par {The information in Fig. \ref{Fig:setup} shows that the peaks can be modeled in at least two different ways: 
\begin{enumerate}\item As half ellipsoids with height $d_0$, or more specifically, as spheroids: ellipses revolved around the axis perpendicular to the plate . \item As cones with height $d_0$. \end{enumerate}}
\noindent {These geometries could produce significantly different results but they are still consistent with Fig. \ref{Fig:setup}.} Strictly speaking, {one should account for the shape of the troughs as well}, but since their contribution is negligible this can be ignored.
\par First we should obtain an estimate of the lateral sizes of the peaks {to make a consistency check: in the model of section \ref{model} each asperity is considered to have a lateral size $\xi$. In the next two paragraphs we will determine which choice of geometry is most consistent with this assumption. The information about the lateral sizes of the peaks} can be extracted from the AFM scans. We have computed the contour of each surface sample at height $d_1$, defined as $d_1\equiv3w$, which is 30.3 nm for the 1600 nm sample. See Fig. \ref{fig:contour1600a} . From the polygon segments of each closed contour the circumferences of the peaks were determined. The associated radii were obtained by assuming circularly shaped bases of the peaks. Typically, high peaks are surrounded by lower peaks, which makes it difficult to distinguish what belongs to the `peak', and what can be considered `normal' roughness. Fig. \ref{fig:contour1600a}  shows that the contours at height $z=0$ cannot be considered circles, whereas the ones at height $z=d_1$ \emph{can}. {For the spheroidal case we can reconstruct their radii at height $z=0$ via the relation
\begin{equation}\label{eq:r_0spheroid}
r_0=\frac{d_0{}r_1}{\sqrt{d_0^2-d_1^2}},
\end{equation}
where $r_1$ represents the (horizontal) radius at $z=d_1$. 
}
\begin{figure*}[!pbt]	\subfigure{\includegraphics[width=0.45\textwidth]{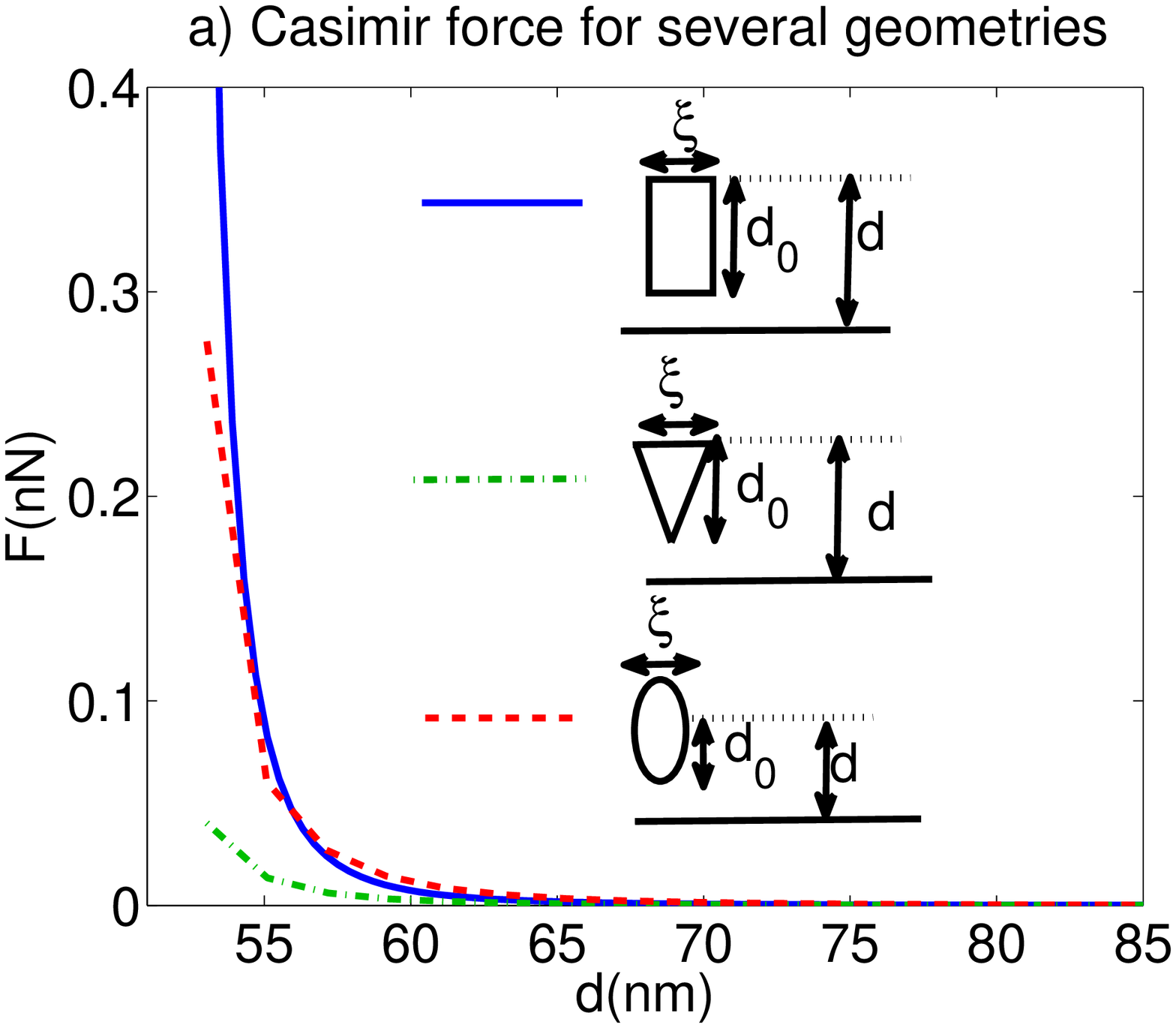}\label{fig:7a}}
\subfigure{\includegraphics[width=0.45\textwidth]{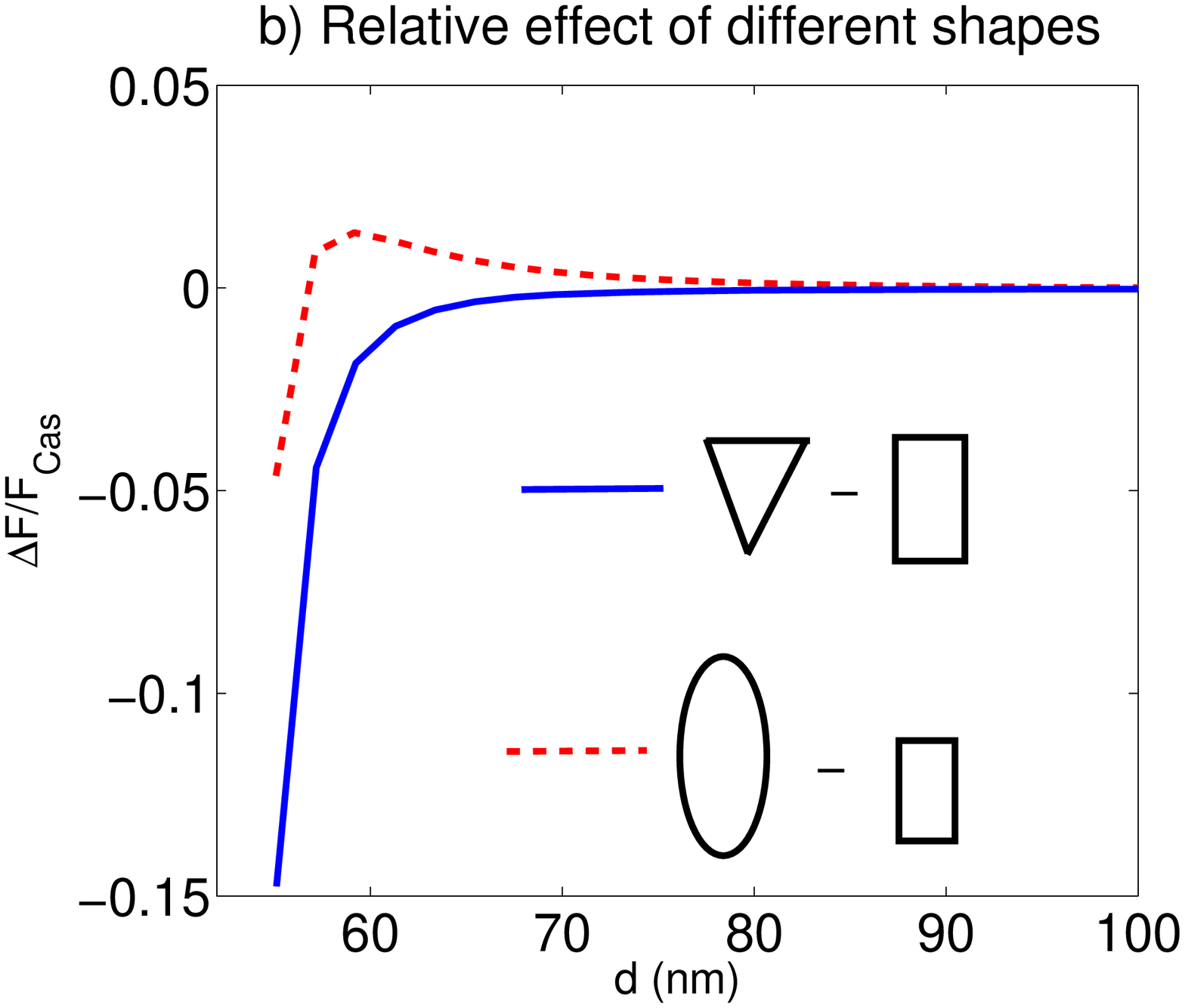}\label{fig:7b}}
	\caption{(color online) Effect of the {shape} of the peaks {for three different geometries. The case of flat peaks was calculated  via the Lifshitz formula. The other two geometries were accounted for by FDTD. In each case, the lateral size is assumed to equal the correlation length $\xi$, and the height was $d_0$.}   {Fig. \subref{fig:7a} shows the absolute force and Fig. \subref{fig:7b} shows the difference $\Delta{}F$ between the FDTD outputs (Eqs. \eqref{eq:F_EP} and \eqref{eq:F_CP}) and the result for flat peaks (Eq. \eqref{eq:Fpp}), relative to} the total Casimir force  $F_{Cas}(d)$  (from Eq. \eqref{modelfinal})}.   
\label{fig:ellipsoid_plate}
	
\end{figure*}
\par With this information we can come to a probability distribution for the radii in the same way as it was done for the heights of the peaks. The probability density function for the radii is shown in Fig. \ref{fig:contour1600b}.  Negative values of $r$ are not allowed, which makes the width distribution $f(r)$ slightly asymmetric, with a skewness of 0.23. Still, $f(r)$ is to good approximation a normal distribution, unlike the height distributions in section \ref{statistics}, where significant deviations from normal distributions were found.
\par This distribution provides an estimate  of the range of values of the lateral sizes of the peaks. The average of this distribution is $44$ nm ($\approx0.9d_0)$, which corresponds to the correlation length, and its standard deviation is 24 nm $\approx0.5d_0$. (See table \ref{tab:parameters}). {Therefore the choice of (half) spheroidal peaks is in this sense a suitable geometry to represent high asperities in this roughness model.}
\par {Similarly the radii for the case of conically shaped peaks at $z=0$ are obtained from the data in Fig. \ref{fig:contour1600a} as follows:
\begin{equation}\label{eq:r_0cone}r_0=\frac{d_0r_1}{d_0-d_1},
\end{equation} 
which means that the distribution in Fig \ref{fig:contour1600b} can still be used, but the variable $r$ must be replaced by $r\rightarrow\sqrt{(d_0+d_1)/(d_0-d_1)}r.$ Consequently, this distribution is much broader than the one for the spheroidal case (Fig. \ref{fig:contour1600b}): the standard deviation is 49.9 nm $\sim{}d_0$. The mean radius is 90 nm in this setup, which deviates considerably from the value of the correlation length listed in table \ref{tab:parameters}. This means that a cone  is not a proper geometry to represent a peak in this model. Modeling the peaks as half spheroids seems preferable in this sense. However, we will still investigate the effect of a conical shape on the Casimir force, so that we might compare it to experimental results in Section \ref{results}.}
\par  The Casimir force between a plate and an ellipsoid {or cone} was calculated numerically. This was done with a finite-difference time-domain (FDTD) \cite{TafloveFDTD} program called Meep. \cite{OskooiRo2010} Recently it was established that FDTD can be used to calculate the Casimir force in arbitrary geometries .\cite{JohnsonPRL2007, Rodriguez2009, McCauley2010} FDTD is a method to numerically solve Maxwell's equations, and its approach  for obtaining the Casimir force is similar to that of Ref. \onlinecite{Lifshitz55}. The main difference, of course, is  that the Green function tensor is obtained numerically in an arbitrary configuration instead of analytically in the parallel plate geometry.
\par We start by separating a conductivity correction factor $C(d)$ from the Lifshitz formula \eqref{PFAcurv}:
\begin{equation}\label{eq:opt_split}
F(d) = C(d)F_{pc}(d)
\end{equation}
where $F_{pc}(d)=\hbar{}c\pi^3R/360d^3$ is the Casimir force between a perfectly conducting plate and sphere in the PFA. Generally there is also a temperature correction factor, but this dependence can be ignored in this separation range.\cite{Milton2004} Note that we have already established the correction factor $C(d)$ from permittivity data obtained via ellipsometry measurements (see Fig. \ref{Fig:ellipsometry}). {We now} perform the calculation of the curvature effect for perfectly conducting bodies and apply the correction $C(d)$ afterwards, as it was done e.g. in Ref. \onlinecite{Lambrecht2000}. {Note that it is assumed here that the effects of the material properties and the shape are independent of each other. Generally, this is not true.\cite{Lambrecht2009} However, at the short separations considered here, the effect of this correlation appears to be small.\cite{Durand2011} This approximation  should suffice to estimate the error due to neglecting the shape of the peaks. In this approximation}  the Casimir force between a dielectric plate and a dielectric spheroid is determined as follows:
\begin{equation}\label{eq:F_EP}
F_{EP}(d) {\approx} C(d)F_{EP,PC}(d),
\end{equation}
{and similarly for the cone-plate geometry
\begin{equation}\label{eq:F_CP}
F_{CP}(d) \approx C(d)F_{CP,PC}(d),
\end{equation}
where $F_{EP, PC}(d)$ and $F_{CP, PC}(d)$ represent the outputs of the FDTD simulation with perfectly conducting bodies}.  The fact that the bodies are perfect conductors and the rotational symmetry of the geometry both reduce the computation time considerably.\cite{McCauley2010}
\par {The result of the FDTD simulations are shown in Fig. \ref{fig:ellipsoid_plate}. They are compared to the force between peaks with flat tips, which is calculated as follows: 
\begin{equation}\label{eq:Fpp}
F_{pp}(d)=-E'(d)\xi^2,
\end{equation}
where $E'(d)$ is determined from the Lifshitz formula  Eq. \eqref{Lifshitz}. This represents the contribution of a single peak in the PFA according to the model outlined in Section \ref{model}. The FDTD calculations were done at separations $d > d_0+$ 2 nm. This is because the FDTD approach requires a surface over which the Maxwell stress tensor is integrated, which in turn requires a buffer between the bodies.\cite{Rodriguez2009, McCauley2010} In the case of curved peaks there is no need to get any closer since the PFA is recovered at short distances. Moreover, the uncertainty in the value of $d_0$ is comparable to 2 nm (see table \ref{tab:parameters}).  It is clear from Fig. \ref{fig:7a} that,  at short separations, the calculation for the spheroidal case is closer to the one for a flat tip than the force between a conically shaped peak and a plate. Fig. \ref{fig:7b} shows that this is also true in a sense relative to the total force of Eq. \eqref{modelfinal}: the maximum effect is almost 5\% in the spheroidal case and about 15 \% in the cone-plate geometry.  In section \ref{results} the relative effects of the shapes of the peaks will be compared to experimental results.}
\par The calculations in this section were performed for the 1600 nm sample only. {This sample has the highest value of the contact distance $d_0$ (See table \ref{tab:parameters}). The experimental uncertainty in the Casimir force \emph{decreases} with $d$ (See section \ref{results}). The effect of the shape of the peaks is most likely to be significant in this case, because the highest value of $d_0$ will not allow lower values of $d$. Moreover, in the approximation of Eqs. \eqref{eq:F_EP} and \eqref{eq:F_CP} $C(d)$ is a monotonically increasing function, so that the total effect will be smaller for samples with a smaller value of $d_0$. In the case of perfect conductors Maxwell's equations are scale-invariant\cite{photonicsbook} and one can use the FDTD outputs for smaller values of $d_0$ as well.}
%
\section{\label{direct_bonding}Direct bonding and surface roughness}

\begin{figure}[!pb]

		\subfigure{\includegraphics[width=0.4\textwidth]{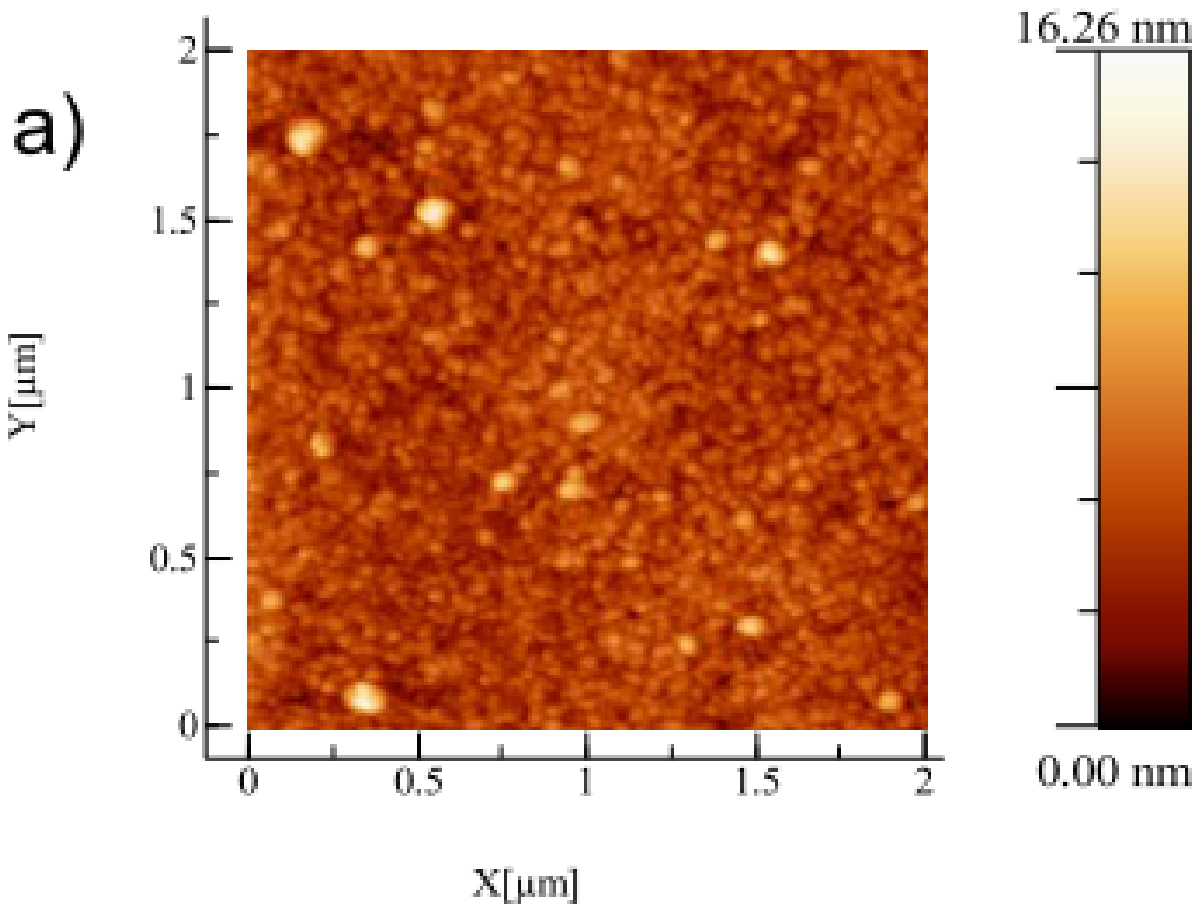}\label{fig:smoothAFMa}}\\
\subfigure{\includegraphics[width=0.4\textwidth]{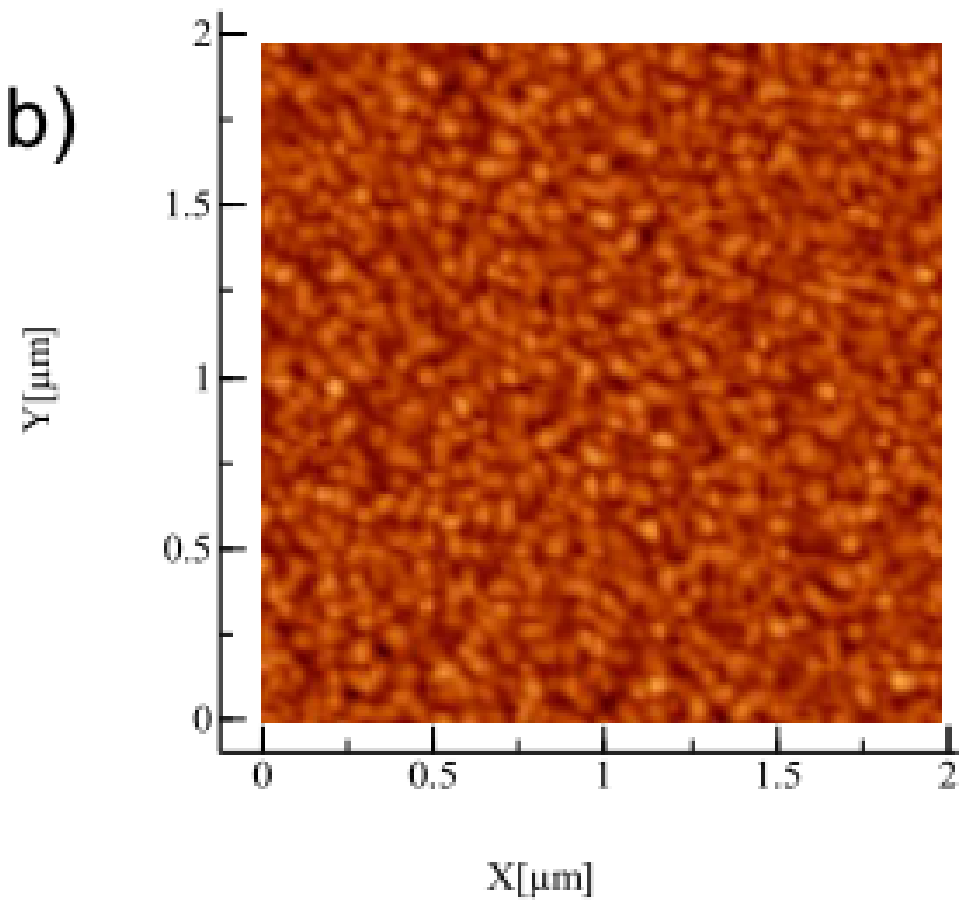}\label{fig:smoothAFMb}}
	\caption{AFM image used for the calculation of the Casimir force in section \ref{direct_bonding}. Same conventions as in Fig. \ref{Fig:AFMscans}. Fig. \subref{fig:smoothAFMa} shows the profile of the sphere and Fig. \subref{fig:smoothAFMb} that of the plate. Both were scanned at a resolution of 512$\times$512 pixels.}
	\label{fig:smoothAFM}
\end{figure}
\begin{figure}[Hb]
	\centering
		\includegraphics[width=0.45\textwidth]{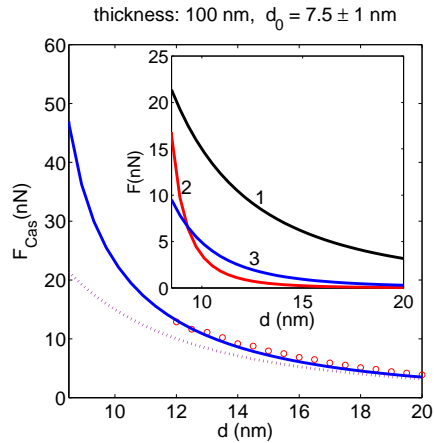}
	\caption{(color online) Calculation of the Casimir force between a 100 nm thick sample with  rms roughness $w=1.3$ nm and a sphere with $w_{sph}=$ 1.8 nm. The three curves in the inset are the three contributions to the solid curve in the main graph:  curve 1 (black) is force without roughness from Eq. \eqref{PFAcurv}, curve 2 (red) is the peaks'contribution  (Eq. \eqref{FPFA}), and curve 3 (blue) is the  perturbative part ({second term of} Eq. \eqref{FPT}).  The contribution of the troughs is always negligible. The dotted line gives the result without roughness effects. The red circles denote experimental points. Measurements {could} not be performed below $d=12$ nm due to jump to contact.}
	\label{fig:Fcas100smooth}
\end{figure}
Since we have established the basics of our approach, we can demonstrate a prediction of the Casimir force in a technologically relevant case: that of relatively low ($<$ 2 nm) rms roughness. In this case the contact distance is also low ($<$ 10 nm) which allows the bodies to move closer to each other, which in turn can give rise to a higher Casimir force.
\par Our studies of the influence of roughness on the Casimir force at close surface proximity, i.e. at separations comparable to $d_0$ are also important for direct bonding technologies. \cite{Haisma2007, Haisma95} Indeed, direct bonding has also become known as van der Waals bonding: Bonding without glue is performed under ambient conditions. Such a bond can only be achieved under strict conditions:\cite{Haisma2007, Haisma95} the geometrical shape of the elements must be optimally congruous; the smoothness of the mechanically finished surface (rms roughness) must be within the subnanometer range; in most cases, the chemical treatment of the surface must be optimum; the physical state of the surface must be defect free; and  the subsurface damage must usually be as small as possible. After annealing and other procedures, \cite{Haisma2007, Haisma95} the direct bond must become  monolithic to guarantee a long life without decohesion of the bonded surfaces.
\par To be more specific: in order to achieve direct bonding  the rms roughness $w$ must be $<$ 2 nm and preferably even $<$ 0.5 nm. \cite{Haisma2007, Haisma95} Such roughness parameter values, at least for the upper roughness limit, have also been  obtained for gold films deposited by electron beam evaporation. \cite{GeorgeAPL2008} In this case, force measurements were only possible down to 12 nm separations due to jump to contact because of capillary forces, while the estimated distance upon contact via height histogram analysis from AFM images was determined to be $d_0=7.5\pm{}1$ nm. \cite{GeorgeAPL2008} In this case of low roughness the Casimir force starts to feel the roughness effect only at separations below 20 nm, as estimations from scattering theory indicated. However, proper analysis of the roughness effect must take into account the contributions of high peaks, especially below 10 nm separations as  $d\sim{}d_0$. These results can be relevant for understanding stiction phenomena under dry conditions (excluding capillary bridge formation) of device components with nanoscale surface roughness, as well as for exploring possibilities of direct bonding phenomena between real surfaces with known optical properties.
\par As jump to contact due to capillary adhesion prevented measurements  at separations below $d=12$ nm, this calculation is a prediction for this range and not a direct comparison to measurements. The experimental data at separations $>$ 12 nm can be reproduced by scattering theory.\cite{GeorgeAPL2008} 
\par The radius of the sphere was 50 $\mu$m, and its rms $w_{sph}=1.8$ nm. while the rms of the plate was $w=1.3$ nm. The AFM scans of the sphere and the plate both had scan sizes of 6$\times$6 $\mu$m$^2$ and 5$\times$5 $\mu$m$^2$, respectively.
\par The results are shown in Fig. \ref{fig:Fcas100smooth}. Near contact, where $d\approx{}d_0$, there are considerable roughness effects: the Lifshitz formula, the ``zeroth order'' perturbative contribution, the black curve no. 1 in the inset, dominates at these short separations, but the contribution of the high peaks (the red curve no. 2 in the inset) is of the same order of magnitude there. The perturbative part (the blue curve no. 3 in the inset) is the smallest contribution, but it cannot be ignored further away from contact where the force was measured. The total Casimir force becomes approximately 46 nN near contact, which is an order of magnitude larger than what has been found for the rougher samples.\cite{PeterRoughness} However, this estimate still needs experimental verification, because presently it is not trivial to measure the force at separations below 10 nm.

\section{\label{results}Results and Discussion}
\begin{figure*}[!ptb]
\centering
\subfigure{\includegraphics[width=0.45\textwidth]{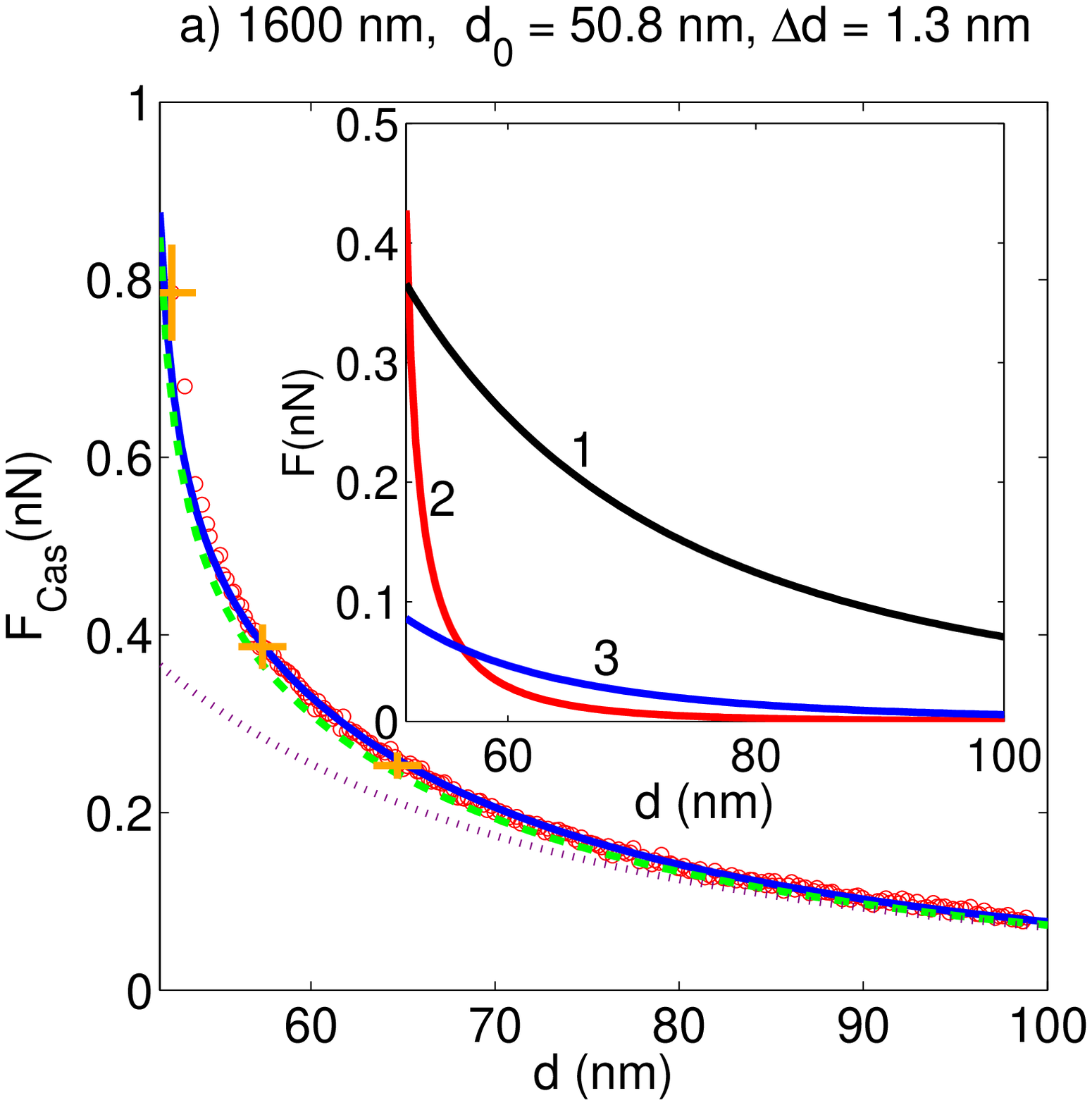}\label{Fig:Fcasa}}
\subfigure{\includegraphics[width=0.45\textwidth]{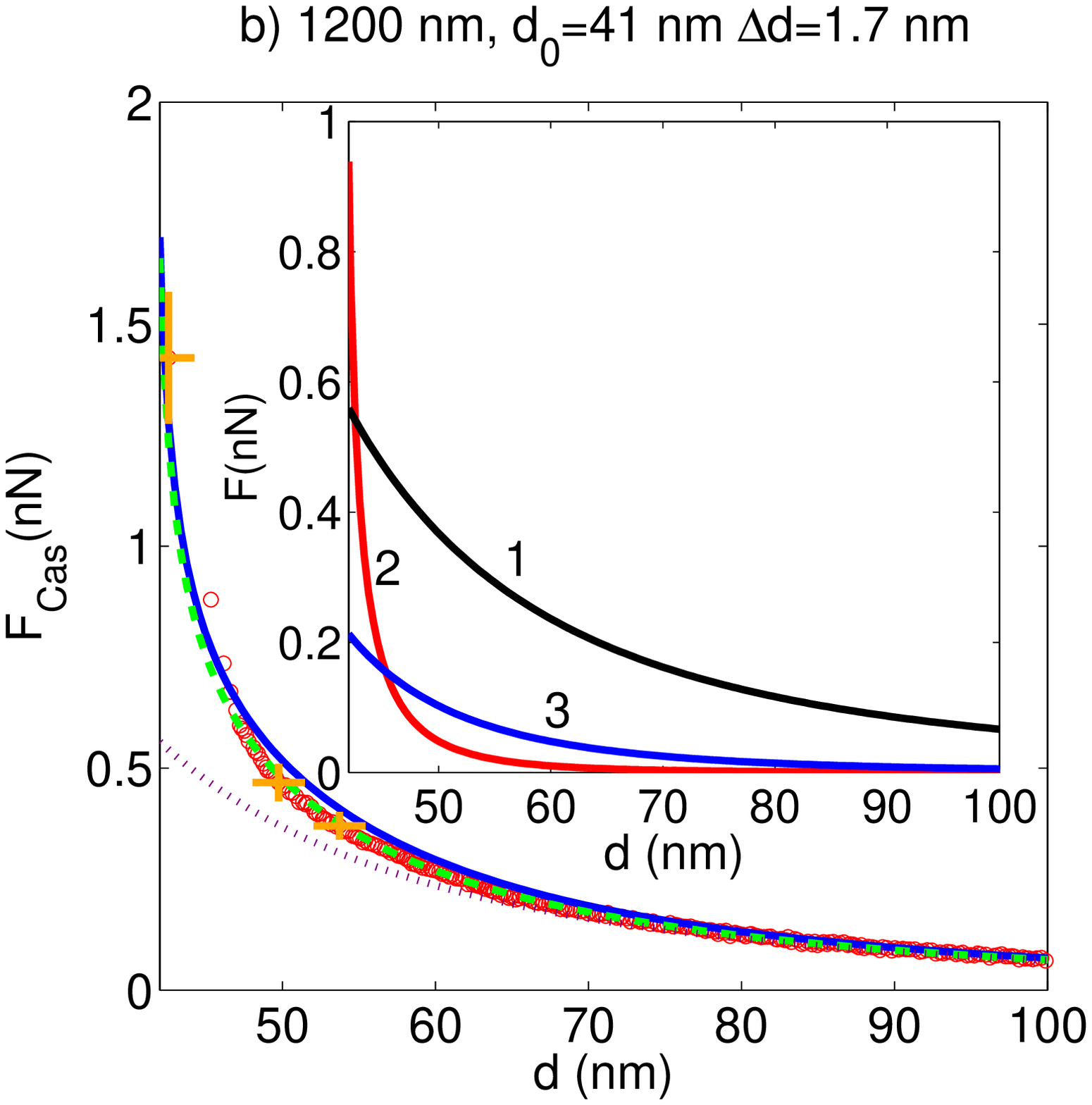}\label{Fig:Fcasb}}\\
\subfigure{\includegraphics[width=0.45\textwidth]{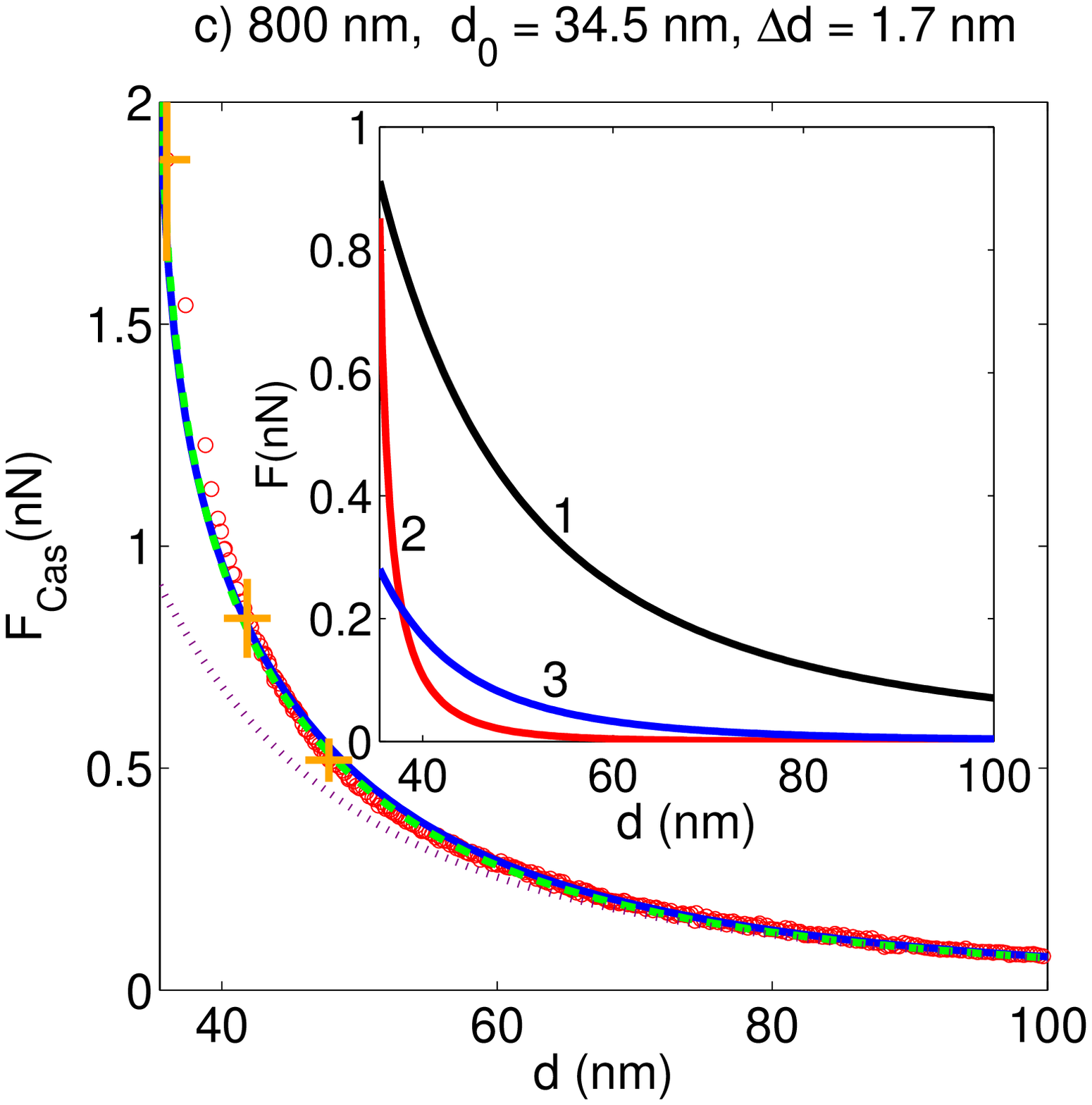}\label{Fig:Fcasc}}
\caption{\label{Fig:Fcas}(color online) Results of the roughness model  (Eq. \eqref{modelfinal}, the solid (blue) line)compared to experimental data (from Ref. \onlinecite{PeterRoughness}, the (red) circles) for three different gold samples.  The (orange) crosses denote the errors in some of the data points. The dashed line is the result of naive application of the PFA via Eq. \eqref{PFArough}. The dotted line is the force without roughness correction. For the insets, the same conventions as in Fig. \ref{fig:Fcas100smooth} apply.}
\end{figure*}
An important question now is: how accurately can we calculate the roughness corrections? The third order term in the Taylor expansion around $z=0$ starting from Eq. \eqref{model2}, $-F'''(d)z^3/3!$, was neglected. In the separation range of interest (20 to 100 nm) the force between smooth surfaces $F(d)$ shows an approximate power law dependence of $d$:\cite{PeterIjmpB2010} $F(d)\sim{}C/d^\alpha$, where $C$ is a constant and the value of the power $\alpha$ depends on the geometry: in the parallel plate setup $\alpha\approx3.5$; in the plate-sphere setup $\alpha\approx2.5$ if $R\gg{}d$. Therefore the estimate of the error due to the use of perturbation theory is given by:
\begin{equation}\label{errperturb}
\Delta{}F_{PT}(d)\approx\gamma\frac{\alpha(\alpha+1)(\alpha+2)}{3!}\Big(\frac{w}{d}\Big)^3F(d),
\end{equation}
where $\gamma$ denotes the skewness of the probability distribution, defined by:
\begin{displaymath}
\gamma\equiv\frac{1}{w^3}\int\limits_{-\infty}^{\infty}z^3f(z)\mathrm{d}z.
\end{displaymath}
The maximum value of $\gamma$ is 1.285 for the 1600 nm sample (see Fig. \ref{Fig:distribution}). In a parallel plate configuration, this leads to $\Delta{}F_{PT}\approx{}18.55(w/d)^3F(d)$. This means that the perturbative contribution has meaning if $d>4w$. The minimum separation distance $d_0$ depends on the scale $L$. It has turned out that even for small $L\approx1\mu$m this condition is usually met.\cite{d0paper} Therefore it is justified to make the important statement that the perturbative contribution has physical meaning up to the point of contact between the interacting bodies.
\par The relative error due to the {assumption that each peak contributes independently} is determined by the condition of its applicability; the distance between the peaks must be sufficiently large: $l(d_1)\gg{}d$. This error is
\begin{equation}\label{PFAerr}
\Delta{}F_{PFA}\approx(d/l)F_{PFA}.
\end{equation}
As we mentioned before, $d_1$ must be chosen in such a way that $l\gg{}d$. One way to do this is {$d_1\equiv3w$} . This definition leads to the values of $l$ listed in table \ref{tab:parameters}. Similarly, we could define $d_1'$ as {$3w$}. However, the contribution of the troughs, $F_{PFA}'(d)$, is always small. It is included only for the sake of generality.
\par In Fig. \ref{Fig:Fcas} the result of our approach (Eq. \eqref{modelfinal}, the continuous (blue) line) is compared to measured force data (from Ref. \onlinecite{PeterRoughness}, the open (red) circles), which were obtained with an AFM setup. The same figure includes the result for a smooth surface (Eq. \eqref{PFAcurv}, the dashed (purple) lines) and that of the PFA for the roughness correction. The latter is given simply by 
\begin{equation}\label{PFArough}
\mathcal{F}_{PFA}(d)=\int\limits_{-d_0'}^{d_0}\mathrm{d}zf(z)F(d-z),
\end{equation}
the results of which are indicated by the dashed (green) lines in Fig. \ref{Fig:Fcas}. Note that this expression is also singular at $d=d_0$. 
%
\par In order to clarify the comparison between measured force data with errors and theoretical predictions as shown in Fig. \ref{Fig:Fcas}, we would like to re-emphasize the distinction between the experimental and statistical error in $d_0$. The values $\Delta{}d$ shown in Fig. \ref{Fig:Fcas} are the \emph{experimental} uncertainties which account only for a \emph{fixed lateral position of the surface profiles with respect to each other}. This is important because the error in $d_0$ dominates the error in the separation distance $d$, and at short separations it also dominates the uncertainty in the Casimir force. This is estimated from the relation
\begin{equation}
\label{Ferror}
\Delta{}F_{Cas}(d)\approx{}F_{Cas}\sqrt{2.5\Bigg(\frac{\Delta{}d_0}{d}\Bigg)^2+\Bigg(\frac{\Delta{}k}{k}\Bigg)^2}.
\end{equation}
The approximate factor $2.5$ can be understood from the fact that $E(d)$ scales approximately as $E(d)\propto{}d^{-2.5}$. \cite{PeterIjmpB2010} The Casimir force was measured with an AFM setup. \cite{PeterRoughness} The relative error in the spring constant $k$ of the cantilever beam is approximately $\Delta{k}/k \approx 3 $\%. The values for $d_0$ and their respective uncertainties have been established from electrostatic calibration and were taken from Ref. \onlinecite{d0paper}. These uncertainties are denoted by error bars through some of the measurements in Fig. \ref{Fig:Fcas}.
\par The insets of Fig. \ref{Fig:Fcas} show the different contributions to the solid lines in the main graphs: curve 1 (black) is force without roughness from Eq. \eqref{PFAcurv} (the ``zeroth order'' perturbative contribution), curve 2 (red) is the peaks' contribution  (Eq. \eqref{FPFA}), and curve 3 (blue) is the second perturbative contribution (Eq.\eqref{FPT}).  The contribution of the troughs, $F_{PFA}'(d)$ (Eq. \eqref{FPFA'}) is always several orders of magnitude smaller than the second smallest  contribution, the second order term in $F_{PT}(d)$ in Eq. \eqref{FPT}. Therefore it is not included in these plots.
\par In each of the three samples in Fig. \ref{Fig:Fcas}, the dashed and the solid line overlap near contact ($d\sim{}d_0$), because the contribution of the peaks is evaluated with the PFA. This contribution decreases very fast with $d$, as the (red) curve labeled as 2 in the inset indicates. This is due to their small area of interaction. This is the dominant contribution near contact for the two roughest samples in Fig. \ref{Fig:Fcasa} and \ref{Fig:Fcasb}. For the other sample, the lower value of $d_0$ allowed the Lifshitz formula to dominate the other contributions. Still, also in this case the peaks contribute considerably near $d=d_0$. A few nm away from contact, the second order perturbative correction (represented by the (blue) curve labeled as number 3 in the inset) starts to become significant. The PFA corresponds to the low wavenumber limit in this contribution.\cite{Genet2003, Lambrecht2005} Therefore it should always dominate the PFA at separations where the contribution of the peaks is negligible. It clearly does for the rougher samples: the solid (blue) line lies above the dashed (green) line in Figs. \ref{Fig:Fcasa}  and \subref{Fig:Fcasb}. For the 800 nm sample (Fig. \ref{Fig:Fcasc}) the contribution from beyond the small wavenumber limit is the smallest; it is barely discernible on the graph.
\par The results of this model are in agreement with measurements for gold samples, unlike perturbation theory, which failed to explain the data. \cite{PeterRoughness} On the other hand, naive application of the PFA via Eq. \eqref{PFArough} also reproduces the the data from Ref. \onlinecite{PeterRoughness} within error. Scattering theory accounts for the non-additivity of the Casimir force and the PFA assumes that it is additive. This indicates that the experiment  in Ref. \onlinecite{PeterRoughness} was not sensitive to the effect of the non-additivity. This is not to say that non-additivity effects are insignificant in general. Indeed, recently significant non-additivity effects have been reported in different contexts, see e.g. Ref. \onlinecite{Chan2008}
\begin{figure*}[!pbt]
	\centering
		\subfigure{\includegraphics[width=0.45\textwidth]{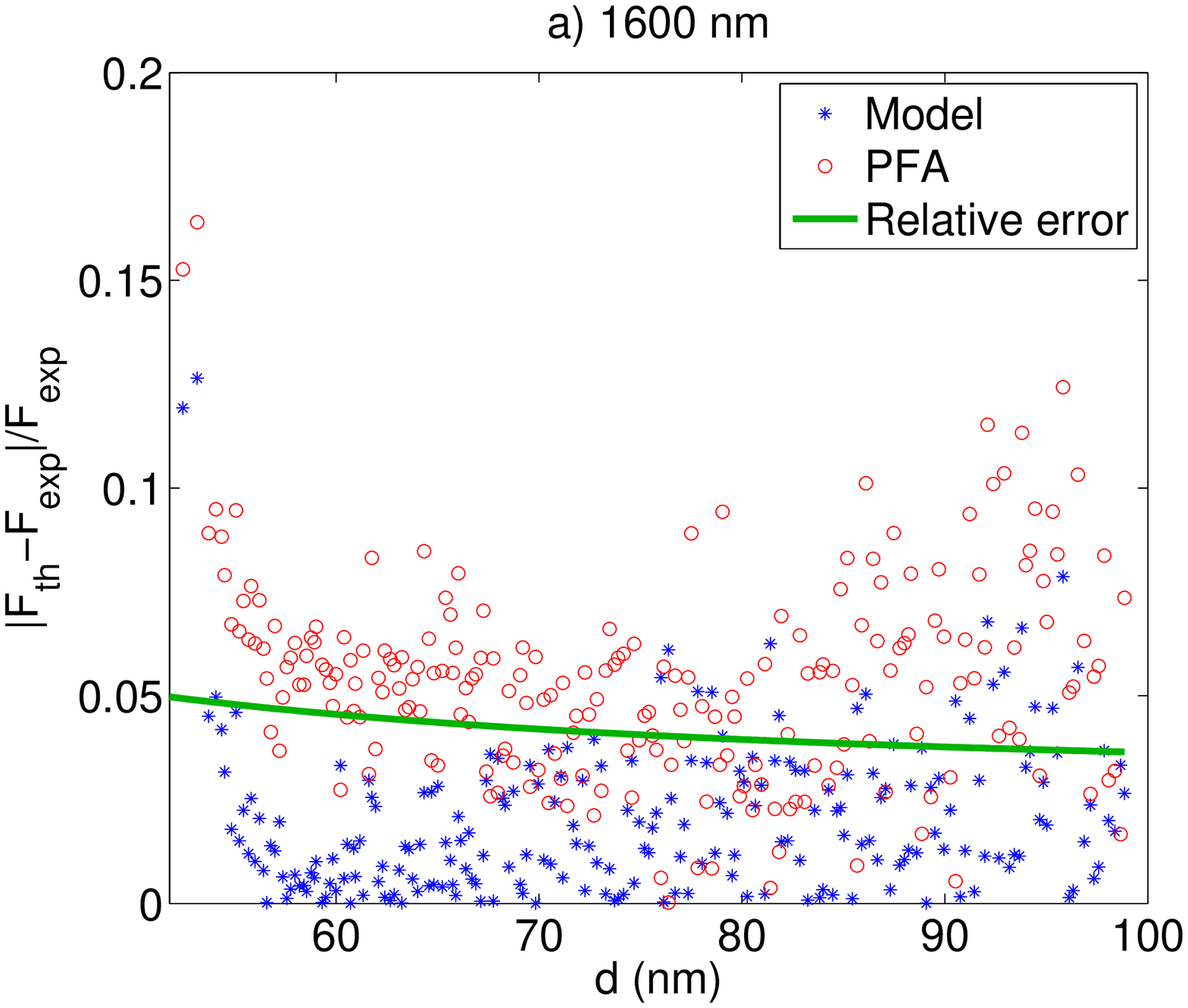}\label{fig:11a}}
	\subfigure{\includegraphics[width=0.45\textwidth]{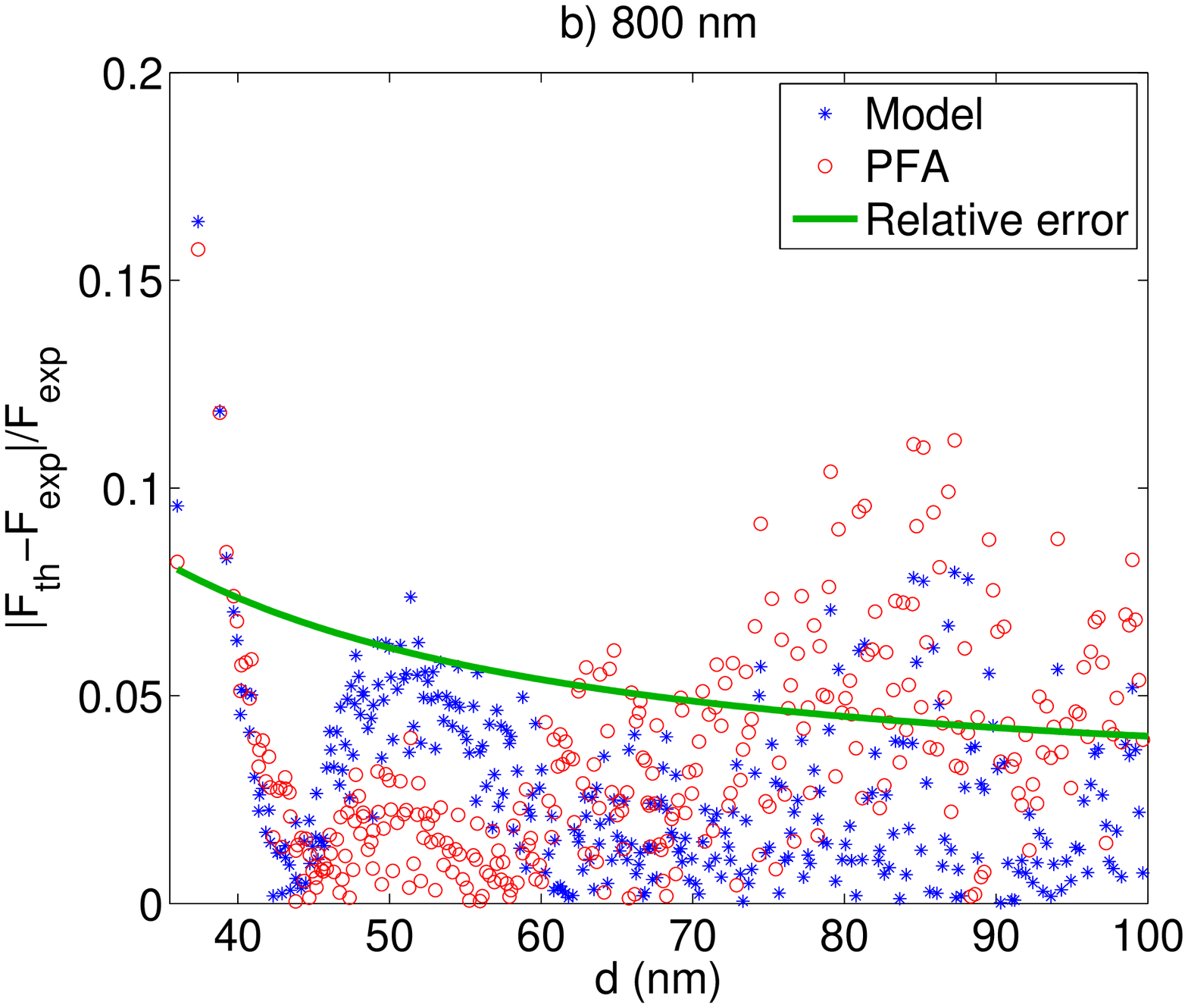}\label{fig:11b}}
	\caption{{(color online) Relative difference between theoretical and experimental results. The solid (green) line represents the relative experimental error from Eq. \eqref{Ferror}. It is assumed that the sphere in the AFM setup has a \emph{fixed} lateral position with respect to the plate. The (red) open circles are comparisons to the naive PFA of Eq. \eqref{PFArough}, and the (blue) asterisks show the difference with the model of Eq. \eqref{modelfinal}.}}
	\label{fig:11}
\end{figure*}\par {The theoretical and experimental results can also be presented in a different way: the absolute value of the relative difference  is plotted in Fig. \ref{fig:11} and compared to the error. The open (red) circles represent the difference with the "naive PFA" of Eq. \eqref{PFArough}, and the blue asterisks show the difference with the model of Eq. \eqref{modelfinal}. The solid (green) line represents the relative error from Eq. \eqref{Ferror}. In Fig. \ref{fig:11a}, which shows the results for the 1600 nm sample, the result of our model , Eq. \eqref{modelfinal}, seems closer to the experimental data than the naive PFA.  However, the difference is less than two standard deviations. This difference is even less pronounced for the 800 nm sample, displayed in Fig. \ref{fig:11b}. In both cases there is a difference of about 15\% at short distances ($d\approx{}d_0$) which exceeds the vertical error. The apparent discrepancy can be accounted for by the horizontal error in Fig \ref{Fig:Fcas}, $\Delta{}d_0$. It should be kept in mind that the force decreases rapidly near contact, so that a small horizontal shift can give rise to a fairly large difference in the vertical direction.}
\par {If the peaks are modeled as half spheroids, the effect of this shape ($\sim$ 5 \%) is still within the experimental error. (See section \ref{curvature}). For conically shaped peaks the effect is 15 \% which is not within the vertical error. This effect is compared to calculations in other geometries, where the value of $d_0$ is exactly the same in each case. Therefore it is independent of the experimental uncertainty in $d_0$, and most likely not responsible for the 15\% difference in Fig \ref{fig:11a}. Moreover, as we found in section \ref{curvature}, conically shaped peaks cannot be reconciled with both the AFM data and the known value of the correlation length. For this reason, cones can be ruled out as a geometry to describe peaks on gold surfaces. However, due to the uncertainty in $d_0$, the measurements of Ref. \onlinecite{PeterRoughness} by themselves do not entirely rule out a 15 \% effect due to the shape of the peaks.}

\section{Conclusions}
We have developed a reliable method to include roughness effects in estimations of the Casimir force at short separations, where perturbation theory fails. Statistical analysis of AFM topography scans has taught us that the surface's height fluctuations can be asymptotically fitted to a Gumbel distribution. We have shown that the contribution of high peaks on a rough surface can be taken into account with the PFA. On the other hand, asperities of height $\sim{}w$ can be evaluated perturbatively. Because the peaks are sufficiently far apart on the scale of the {separation}, their contributions to the Casimir force are statistically independent.
\par It has been established that the peaks contribute significantly to the Casimir force, particularly near contact where $d\approx{}d_0$. The high peaks not only dominate the force, but they also shift the minimum separation distance from $0$ to $d_0$. To a large extent this gives rise to the scaling of the force observed experimentally: the shift of the singularity in the Lifshitz formula makes both experimental and theoretical curves singular at $d\approx{}d_0$ and unphysical below $d_0$. The inclusion of contact between the bodies appears to be a crucial aspect of the roughness correction to the Casimir force.
\par We have presented detailed calculations of the influence of the curvature of the peaks by modeling them as half spheroids, but this has a marginal effect on the force as a whole. The reason for this is that their contribution is significant only near contact (where $d-d_0\ll\xi$), and decreases rapidly with $d$ due to their small area of interaction. In this near contact limit the PFA is valid, so that we can neglect the curvature of the peaks. {On the other hand, modeling the peaks as cones cannot reproduce the correlation length from the information that the AFM data provides about the lateral sizes of the peaks. It can be concluded that cones are not a proper geometry to describe peaks on gold surfaces. Moreover, it produces an effect that does not seem to be well supported by experiment, even though it cannot be entirely ruled out either.}
\par We have calculated the Casimir force between relatively  smooth  surfaces,  which  is potentially  useful for direct bonding applications.   It was found that the Casimir force is an order of magnitude higher than the force between rougher surfaces, because the lower value of the contact distance allowed lower separations.  Possibly, higher Casimir forces could be achieved between congruous bodies.  In such a case, this approach for the
roughness correction could be combined with numerical methods (e.g.  FDTD \cite{Rodriguez2009} ) to account for the geometry of the system. Such a calculation would be computationally challenging, because it involves multiple scales.
\par It has also turned out that naive application of the PFA described by Eq. \eqref{PFArough} gives a result close to that of our approach and hence is also in good agreement with the experiment. Perturbation theory accounts for the non-additivity of the Casimir force whereas the PFA assumes that it is additive. Apparently, the experiment in Ref \onlinecite{PeterRoughness} was not sensitive to the effect of non-additivity.
\par 
Notably, the significance of the role of the peaks in the roughness correction can also be of interest for problems of capillary adhesion between surfaces,\cite{DelRio2005} including wet environments.\cite{DelRio2007, PeterCapillary, Persson2008}

\begin{acknowledgments}
We acknowledge useful discussions with M. Kardar,  P. J. van Zwol and B. J. Hoenders. The authors benefited from exchange of ideas within the ESF Research Network CASIMIR.
\end{acknowledgments}

\bibliography{Casimir}

\end{document}